\documentclass{iopart}
\pdfoutput=1
\usepackage{iopams}  

\usepackage{graphicx}
\usepackage{xcolor}
\usepackage{aas_macros}
\usepackage{hyperref}
\usepackage{amstext}

\def\Omg{\hat\Omega}
\def\muu{\mu_{\rm u}}
\def\g{\gamma}
\def\G{\Gamma}
\def\d{{\rm d}}

\def\be{\begin{equation}}
\def\ee{\end{equation}}
\def\bea{\begin{eqnarray}}
\def\eea{\end{eqnarray}}
\def\bi{\begin{itemize}}
\def\ei{\end{itemize}}
\def\ben{\begin{enumerate}}
\def\een{\end{enumerate}}
\def\bnp{\numparts}
\def\enp{\endnumparts}
\def\implies{\Rightarrow}
 
\newcommand{\Srefs}[1]{Sections~\ref{#1}}

\usepackage{etoolbox}
\makeatletter
\newcommand{\mainmatter}{%
  \setcounter{footnote}{2}%
  \patchcmd{\@makefntext}{\fnsymbol}{\arabic}{}{}%
  \patchcmd{\@thefnmark}{\fnsymbol}{\arabic}{}{}%
  \def\@makefnmark{\textsuperscript{\arabic{footnote}}}%
}
\makeatother

\begin{document}

\title[Hellings and Downs FAQ sheet]{Answers to frequently asked 
questions about the pulsar timing array Hellings and Downs curve}

\author{J D Romano$^1$
and B Allen$^2$}

\address{$^1$
Department of Physics and Astronomy,
University of Texas Rio Grande Valley,
One West University Boulevard,
Brownsville, TX 78520, USA} 

\address{$^2$ 
Max Planck Institute for Gravitational Physics
(Albert Einstein Institute), Leibniz Universit\"{a}t Hannover, 
Callinstrasse 38, D-30167 Hannover, Germany}

\eads{\mailto{joseph.romano@utrgv.edu}, 
\mailto{bruce.allen@aei.mpg.de}}
\vspace{10pt}
\begin{indented}
\item[]Jan 2024
\end{indented}

\begin{abstract}
We answer frequently asked questions (FAQs) about the
Hellings and Downs correlation curve---the ``smoking-gun" signature
that pulsar timing arrays (PTAs) have detected gravitational waves
(GWs).  Many of these questions arise from inadvertently applying
intuition about the effects of GWs on LIGO-like detectors to the 
case of pulsar timing, where not all of it applies.
This is because Earth-based detectors, like LIGO and Virgo,
have arms that are short (km scale) 
compared to the wavelengths of
the GWs that they detect ($\approx 10^2$\textendash$10^4$~km).  
In contrast, PTAs respond to GWs whose wavelengths (tens of light-years) 
are much shorter than their arms
(a typical PTA pulsar is hundreds to thousands of light-years from Earth).  
To demonstrate this, we calculate the 
time delay induced by a passing GW along an Earth-pulsar baseline (a
``one-arm, one-way" detector)
and compare it in the ``short-arm" (LIGO-like) and ``long-arm" (PTA)
limits.  This provides qualitative and quantitative answers to many
questions about the Hellings and Downs curve.  The resulting
FAQ sheet should help in understanding the ``evidence for GWs"
recently announced by several PTA collaborations.
\end{abstract}

%
\vspace{2pc}

\mainmatter

\section{Introduction}
\label{s:intro}

Four pulsar timing array (PTA) collaborations recently announced the
results of their latest searches for correlated low-frequency
($\sim\!10^{-9}~{\rm Hz}$) gravitational waves (GWs).  The
findings span the range from 
``weak evidence" for GWs for the (Australian) Parkes Pulsar Timing Array~\cite{PPTA-GWB:2023},
``some evidence" for GWs for the Chinese Pulsar Timing Array~\cite{CPTA-GWB:2023}, 
``evidence/marginal evidence" for GWs for the European Pulsar Timing Array~\cite{EPTA-GWB:2023}, 
and ``compelling evidence" for GWs for the North American Nanohertz Observatory for Gravitational
Waves (NANOGrav)~\cite{NG-GWB:2023}.

These observations can be challenging to understand, in part because
the observed signal in the PTA band is unlike the short duration
($\sim\!0.25~{\rm s}$) GW ``chirp" signal that LIGO observed on 14 Sep
2015, from the final inspiral and merger of two $\sim\!30 M_\odot$
black holes.  Instead of short (1~s) audio frequency (100~Hz)
deterministic GW signals, PTAs search for persistent (signals with
correlation times $\lesssim 500$~yr~\cite[Fig.~9]{AllenHDVariance}
and which are in band for $\sim\!10^6$~yr), low-frequency (nHz) 
stochastic GW signals.  These
are too feeble to observe directly, but should produce correlated
perturbations in the arrival times of pulses from an array of galactic
millisecond pulsars.
  
Although the current PTA observations are not statistically significant enough to
unambiguously claim a detection or to identify a source, a
credible candidate is the superposition of GWs from hundreds of
thousands of supermassive ($\sim\!10^9 M_\odot$) black holes (SMBH)
orbiting one another in the centers of merging galaxies.
While any one of those binaries would produce a deterministic
almost-constant-frequency signal, the incoherent sum of the GWs
from many such sources creates a stochastic confusion noise: a ``hubbub" 
or ``hum" of GWs.

PTAs search for correlations that are predicted to follow the
so-called ``Hellings and Downs" curve.  This correlation curve is
named after Ron Hellings and George Downs, who first derived it in
1983~\cite{HD}.  The idea is that pulsars act as high-quality clocks,
emitting radio pulses that arrive at Earth\footnote{Our treatment is
simplified: it assumes that the Earth is isolated, and that it is not 
orbiting the Sun, orbited by the Moon, or affected by other planets.  
In practice, these and other deterministic effects are removed
by converting all pulse arrival times to pulse arrival times
at the center of gravity of the solar system, which is called the
Solar System Barycenter (SSB).  The SSB is located near the surface of
the Sun, about 8 light-minutes from Earth.  On the scale of the GWs
of interest, this is close enough to the Earth that we do not make a
distinction here.}  with very predictable and regular arrival times.
GWs induce slight perturbations in these arrival times, and these
perturbations (called {\it timing residuals}) are correlated between
different pulsars in a very specific fashion.

\begin{figure}[htbp]
\begin{center}
\includegraphics[width=.65\textwidth]{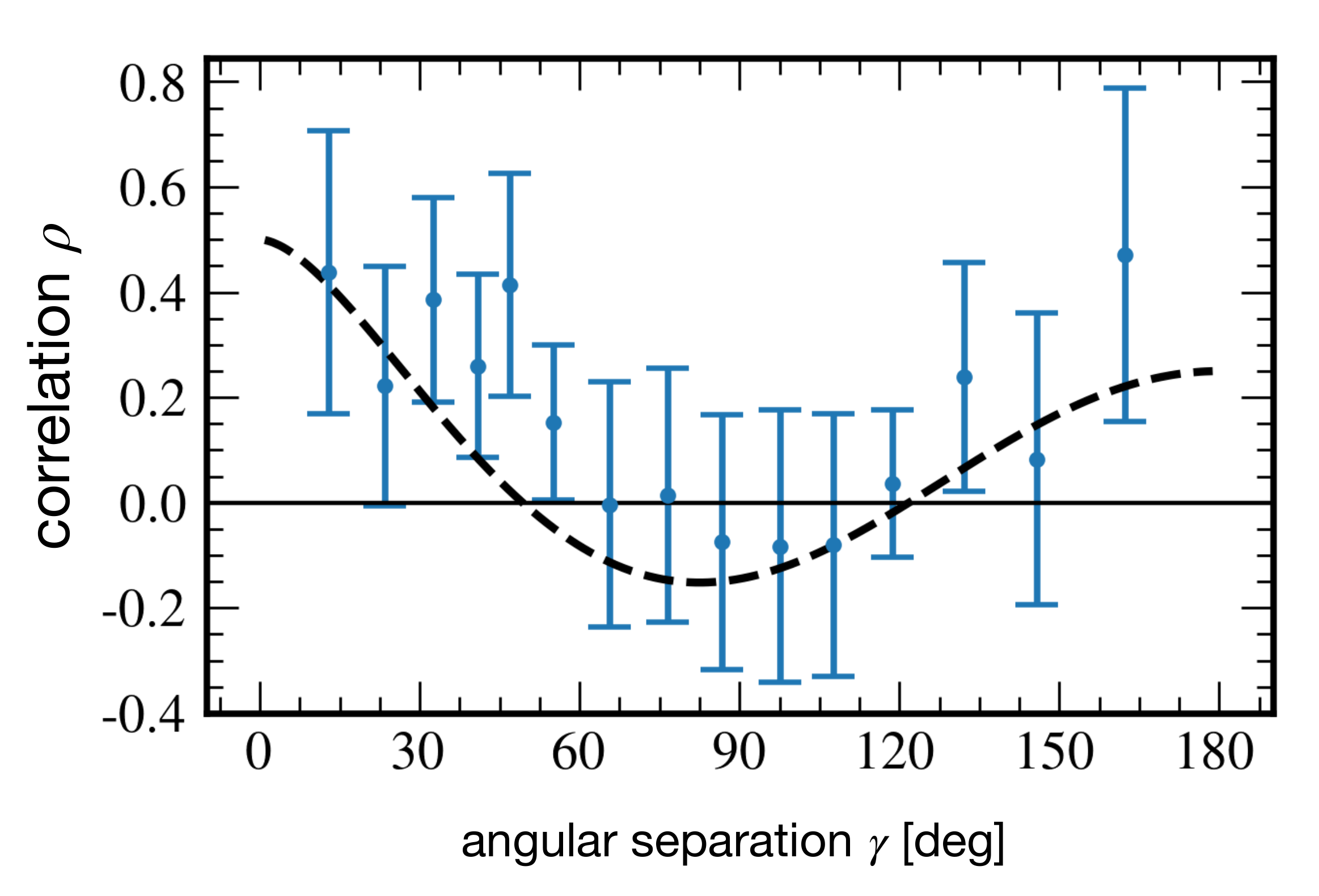}
\caption{The spatial correlations observed in the pulsar
    timing residuals for the NANOGrav 15-year
    dataset~\cite{NG-GWB:2023} are shown in blue; the Hellings
    and Downs curve/prediction is shown in black.  The blue points and error
    bars are optimally weighted averages of approximately 150
    pulsar-pair correlations in each angular separation bin, which
    take into account covariances between the correlations induced by the
    GWB itself.  (Reproduced from Figure~2c of~\cite{NG-GWB:2023}.)}
\label{f:HD-NG}
\end{center}
\end{figure}

The Hellings and Downs curve is a plot of the {\it expected} or {\it
  average} correlation in the timing residuals from a pair of pulsars,
as a function of their angular separation on the sky as seen from
Earth.  This curve is a prediction of general relativity (GR) for a
gravitational-wave background (GWB) that is {\it unpolarized} and {\it
  isotropic}.  More precisely, the Hellings and Downs curve is the
expected value of the pulsar-pair correlations averaged over all
pulsar pairs having the same angular separation and over GW sources
having independent, random phases.  Later, we will discuss several
different interpretations for the curve, depending upon the type of
averaging which is employed.  Figure~\ref{f:HD-NG} shows how the
timing residuals in NANOGrav's 15-year dataset~\cite{NG-GWB:2023}
appear to match this predicted Hellings and Downs curve.

\subsection{Frequently asked questions about the Hellings and Downs
curve and the Hellings and Downs correlation}
\label{s:FAQs}

When giving talks about the recent PTA observations, the authors have
repeatedly been asked certain questions about the properties of the
Hellings and Downs correlation curve.  The most common questions vary
a bit, but can be summarized as:

\ben
\item[Q1:]
Why is the Hellings and Downs curve in Figure~\ref{f:HD-NG} 
normalized to 1/2 at zero angular separation, but is normalized 
to $1/3$ in other papers? 
Does it matter?

\item[Q2:]
Why does the Hellings and Downs curve have different values at
$0^\circ$ and $180^\circ$ if the quadrupolar deformation of space
produced by a passing GW affects two test masses $180^\circ$
apart in exactly the same way?

\item[Q3:] 
Why is the value of the Hellings and
Downs curve for two pulsars separated by $180^\circ$ 
exactly half that for $0^\circ$ angular separation?

\item[Q4:]
Why is the most negative value of the Hellings and Downs 
curve not at $90^\circ$?

\item[Q5:]
Why is the Hellings and Downs curve frequency independent,
whereas overlap functions for Earth-based interferometers
are frequency dependent?

\item[Q6:]
Does recovery of the Hellings and Downs curve imply that
the GWB is isotropic?

\item[Q7:] In the distant future, when PTA observations are carried
  out with larger and more sensitive telescopes, more pulsars, and
  much longer observation periods, will we {\it exactly} recover the
  Hellings and Downs curve?

\item[Q8:]
Assuming noise-free observations, are the fluctuations
away from the expected Hellings and Downs curve only due to 
the pulsar-term contributions to the timing residual response?

\item[Q9:]
Why does the Hellings and Downs recovery
plot shown in Figure~\ref{f:HD-NG} have error bars that are
much larger than one would expect for the reported $3-4\sigma$
significance?

\item[Q10:]
If a deviation from the Hellings and Downs curve is detected, what
does that mean?

\end{enumerate}

We believe that many of these questions arise because much of our
intuition regarding the response of a detector to a passing GW was
developed for Earth-based GW detectors such as LIGO and Virgo. Not all
of it carries over to the case of PTAs.  This is because the distance
between the Earth and a pulsar (one ``arm" of a PTA detector) is
typically hundreds to thousands of times longer than the wavelengths
($\lambda \approx 10^0-10^1 \, \text{light-years}$) of the GWs that
PTAs are trying to detect.  In contrast, Earth-based interferometers
have km-long arms that are much shorter than the wavelengths ($\lambda
\approx 10^2$\textendash$10^4$~km) of the GWs that these
detectors are sensitive to.  In summary: PTA detectors operate in the
``long-arm" (or short-wavelength) limit, whereas Earth-based laser
interferometers operate in the ``short-arm" (or long-wavelength)
limit.

The goal of this paper is to provide qualitative and quantitative
answers to the questions above, highlighting the differences between
PTAs and LIGO-like detectors as appropriate.

Note that, as suggested by Q7, we make a distinction between (a) the
Hellings and Downs {\it curve}, which is the black curve shown in
Figure~\ref{f:HD-NG} and (b) the observed Hellings and Downs
bin-averaged {\it correlation}, shown by the blue points and error bars
in the same figure, which might or might not follow this curve.

\subsection{Outline}
\label{s:outline}

The remainder of this paper is organized as follows.
There are two main parts: 
Part I consists of 
\Srefs{s:single_pulsar} and \ref{s:correlated_response}, 
which provide the background information
needed to derive the Hellings and Downs 
curve. This includes the calculation of the exact 
response of a single Earth-pulsar baseline to a 
passing GW, and two different averaging procedures
used to obtain the expected correlation for pairs of
Earth-pulsar baselines.
Part II consists of 
\Srefs{s:answers} and \ref{s:summary}, 
which provide the answers to the FAQs listed above 
and a brief summary of the key points of the paper.

Readers who are already familiar with pulsar timing and the 
definition of the Hellings and Downs correlation could 
skip \Srefs{s:single_pulsar} and \ref{s:correlated_response} 
and jump directly to \Sref{s:answers}.
However, we advise against this for two reasons:
(i) In \Srefs{s:redshift_response} and \ref{s:antenna_patterns},
we compare the response of ``long-arm" PTA pulsars to 
that of ``short-arm" LIGO-like detectors, which we believe
is the source of many of the misconceptions regarding the 
Hellings and Downs curve.
(ii) Our discussion of pulsar averaging in \Sref{s:pulsar_averaging} 
will be new to many, and the concepts introduced 
there are important for understanding what follows.

To make this article more accessible, we take a pedagogical approach.
We avoid jargon, and either give ``first-principles" explanations or
provide pointers to papers containing these.  We also suggest
exercises for the reader. While no solutions are given, for the more
involved exercises we point to published papers that contain these solutions.

From here forward, we use units in which the propagation speed $c$ of
light and of GWs is set equal to one.  This means that both time and
distance are measured in units of seconds or years as appropriate.
More precisely, a distance is given by the time that it takes light to
traverse that distance.  For example, a meter stick is approximately
3~nanosec long, the LIGO detectors' arms are about 10 microseconds long, the
wavelengths of the GWs that PTAs are sensitive to are typically
10~years long, and typical pulsars are thousands of years from Earth.

We also adopt the Einstein summation convention for repeated spatial
indices such as $i$ and $j$.

\section{Single pulsar response}
\label{s:single_pulsar}

\subsection{Single GW point source}
\label{s:plane_wave}

\begin{figure}[htbp]
\begin{center}
\includegraphics[width=.6\textwidth]{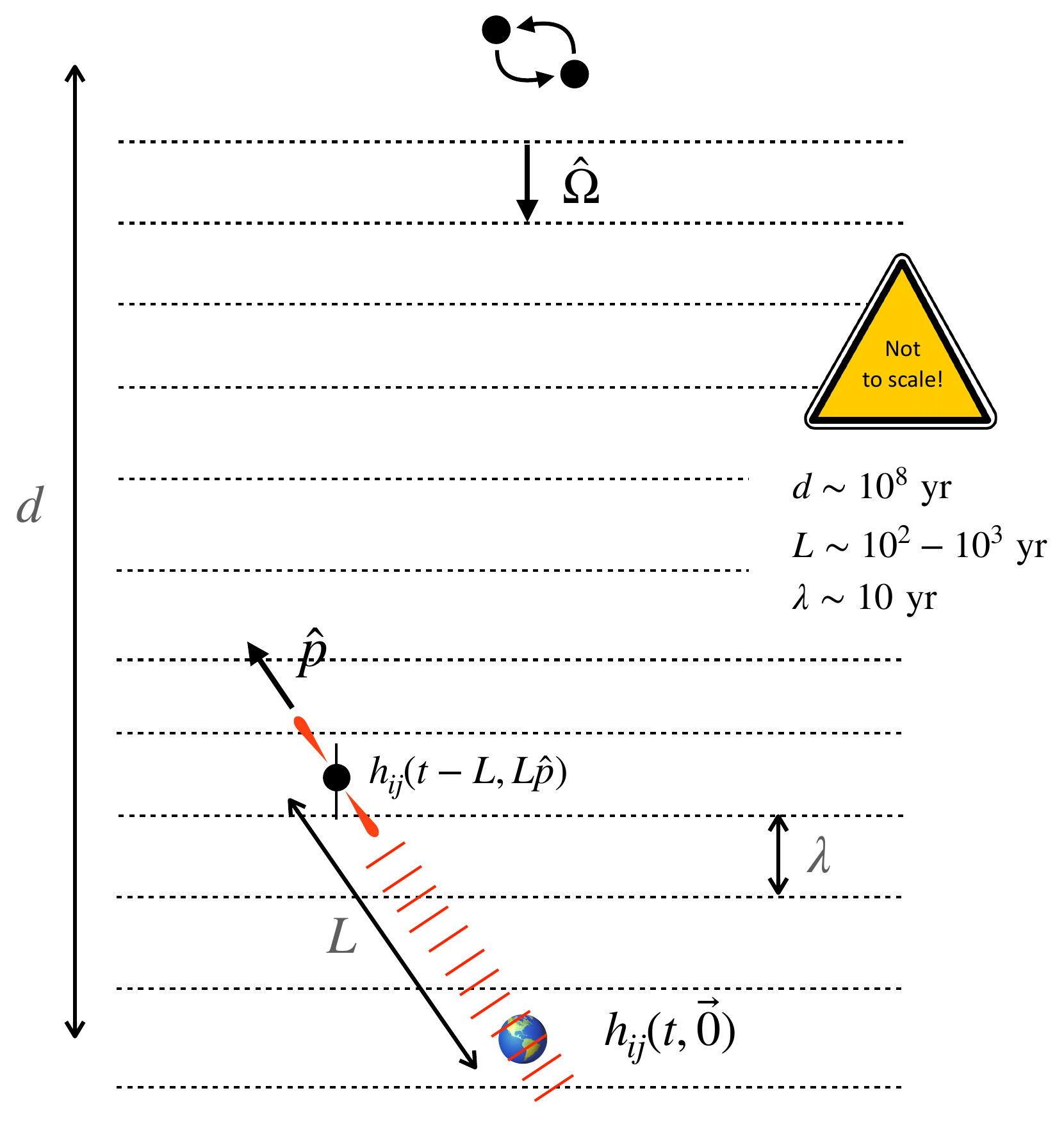}
\caption{Geometry for the single pulsar analysis.
The Earth is located at the origin of a rectangular
Cartesian coordinate system. The components
$h_{ij}(t,\vec x)$ of the metric perturbations are functions
of time and of the spatial coordinates $\vec x$.
A single point source of GWs is located a distance 
$d$ from Earth in direction $-\Omg$ on the sky.
A galactic pulsar is located a distance $L$ 
from Earth in direction $\hat p$.
The distances $\lambda$, $L$ and $d$ are related by 
$\lambda \ll L \ll d$, where $\lambda$ is the 
characteristic wavelength of the GWs that 
PTAs search for (of order 10 light-years).}
\label{f:one_pulsar_geom}
\end{center}
\end{figure}

We begin by describing the GWs emitted by a single very distant point
source, and consider their effect on light (i.e., a radio pulse)
propagating along an Earth-pulsar baseline.  For example, imagine that
the GWs are produced by an orbiting pair of SMBHs in the centers of two
merging galaxies,
$\sim\!10^8$~years from Earth.  The SMBHs are separated by a
distance of a few years, so from the perspective of Earth are a point
source.  For such a large distance from the source, the associated
metric perturbations $h_{ij}(t,\vec x)$ are solutions of the vacuum
wave equation and are well approximated in the neighborhood of the
Earth and pulsar by plane waves propagating in direction $\Omg$, where
$-\hat\Omega$ is the direction from Earth to the SMBH pair
(see \Fref{f:one_pulsar_geom}).

\begin{figure}[htbp]
\begin{center}
\includegraphics[width=.7\textwidth]{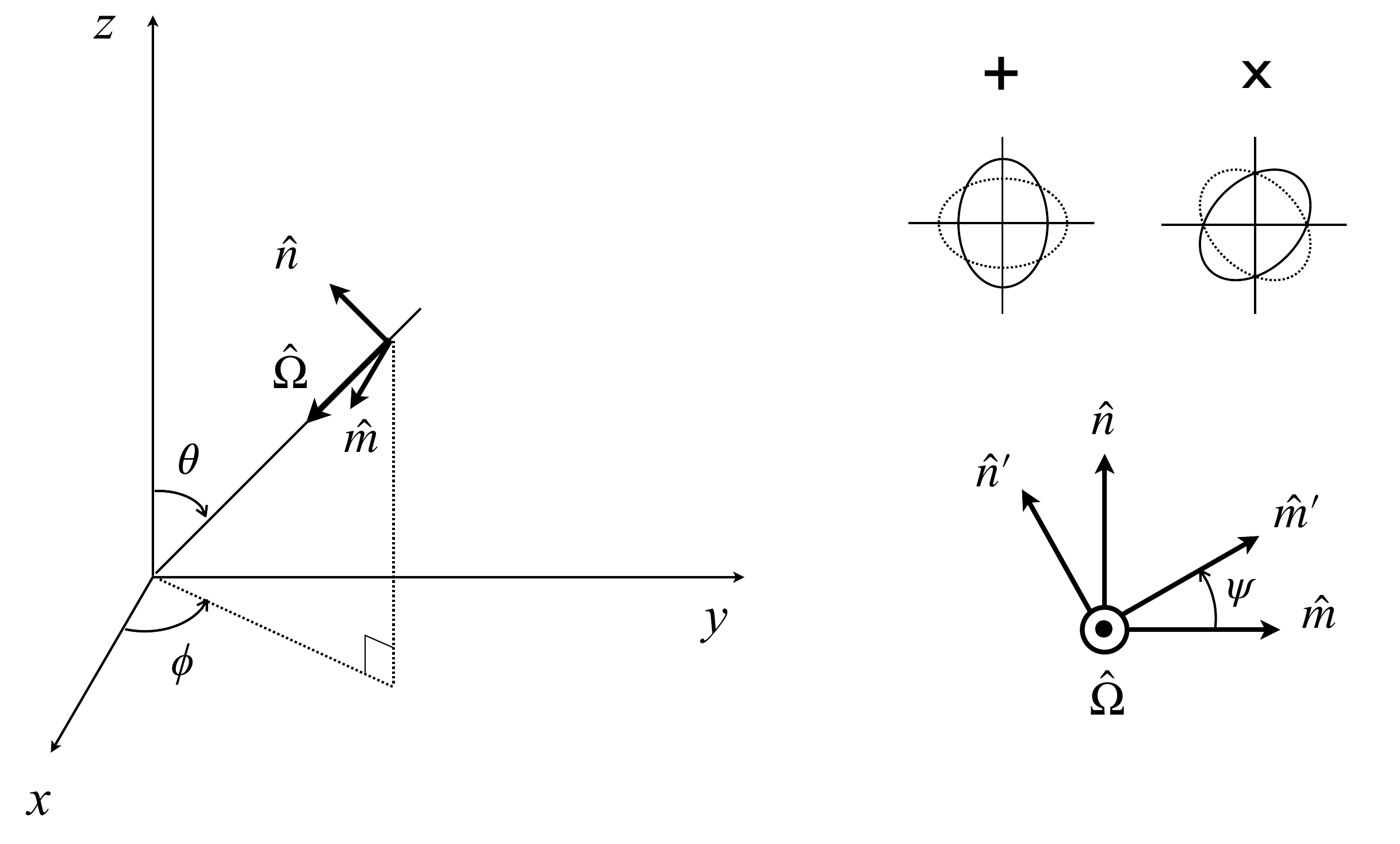}   
\caption{Left panel: Coordinate system and unit vectors used in the
  plane wave expansion.  Right panel (top): two orthogonal
  polarizations of a GW.  A circular ring of freely falling test
  particles is placed at rest in the plane orthogonal to the
  propagation direction $\Omg$ of the wave. These are periodically
  deformed into ellipses, as space is ``squeezed" and ``stretched" 
  (here very exaggerated) by
  the passing of the GW.  Right panel (bottom): an alternative basis
  of two orthogonal polarization vectors may be obtained by rotating $\hat m$
  and $\hat n$ through angle $\psi$ to obtain new vectors $\hat m'$
  and $\hat n'$.} 
\label{f:plane_wave_polarizations}
\end{center}
\end{figure}

In transverse-traceless-synchronous coordinates
$(t,\vec x)\equiv (t,x^i)$, where $i=1,2,3$, we have%
\footnote{It is immediate from the functional form
of $h^A(t-\hat\Omega\cdot\vec x)$, and the definitions of the polarization 
tensors $e^+_{ij}(\Omg)$, $e^\times_{ij}(\Omg)$ 
[see \eref{e:e+}, \eref{e:ex}] in terms of the 
unit vectors $\hat\Omega$, $\hat m$, $\hat n$ 
[see \eref{e:Omg}, \eref{e:m}, \eref{e:n}] that the 
metric perturbations 
$h_{ij}(t,\vec x)$ satisfy the vacuum wave equation,
$(\partial^2/\partial t^2 - \nabla^2)h_{ij}=0$,
are transverse $h_{ij}\hat\Omega^j=0$, and traceless
$h_{ij}\delta^{ij}=0$. The coordinates are called ``synchronous'' because
the metric perturbations are purely spatial: their time components vanish.}
\bnp
\bea
h_{ij}(t,\vec x) 
&= \sum_{A=+,\times} h^A(t-\Omg\cdot\vec x)e^A_{ij}(\Omg)
\label{e:h_ij_real}
\\
&= \Re\left[h(t-\Omg\cdot \vec x) e^*_{ij}(\Omg)\right]\,,
\label{e:h_ij_comp}
\eea
\enp
where
\be
h(u) \equiv h^+(u) +ih ^\times(u)\,,
\quad
e_{ij}(\Omg)\equiv e^+_{ij}(\Omg)+ie^\times_{ij}(\Omg)
\label{e:h_e_ij_comp}
\ee
are complex combinations of the arbitrary real functions
$h^+(u)$, $h^\times(u)$ and real polarization tensors
$e^+_{ij}(\Omg)$, $e^\times_{ij}(\Omg)$ defined by
\bnp
\bea
e^+_{ij}(\Omg) = \hat m_i \hat m_j - \hat n_i \hat n_j\,,
\label{e:e+}
\\
e^\times_{ij}(\Omg) = \hat m_i \hat n_j + \hat n_i \hat m_j\,.
\label{e:ex}
\eea
\enp
Here, $\hat m$ and $\hat n$ are any two 
orthogonal unit vectors in the plane perpendicular to $\Omg$.
For stochastic background analyses, a common convention which we adopt here is
\bnp
\bea
\Omg\equiv
-\hat x\,\sin\theta\cos\phi
-\hat y\,\sin\theta\sin\phi
-\hat z\,\cos\theta = -\hat r\,,
\label{e:Omg}
\\
\hat m\equiv
\hat x\,\sin\phi
-\hat y\,\cos\phi = -\hat\phi\,,
\label{e:m}
\\
\hat n\equiv
-\hat x\,\cos\theta\cos\phi
-\hat y\,\cos\theta\sin\phi
+\hat z\,\sin\theta = -\hat\theta\,,
\label{e:n}
\eea
\enp
where $\hat r$, $\hat\theta$, $\hat\phi$ are the standard
spherical polar coordinate unit vectors, which are normal and tangent to the two-sphere.
The vectors $\Omg$, $\hat m$, and $\hat n$ are
shown in 
\Fref{f:plane_wave_polarizations}.
%
The reader should be aware that in the literature there are two
commonly-used mappings between the wave propagation direction and
spherical polar coordinates $\theta, \phi$.  In this work, as in
\cite{Romano2017,Sesana_2013} the minus signs that appear in the
mapping \eref{e:Omg} mean that $\theta, \phi$ are the sky coordinates
of the GW \emph{source}.  In the other papers that we cite~\cite{ABCPS,
  AllenHDVariance,AllenRomano,allen2024pulsar,PhysRevD.90.082001},
$\theta, \phi$ are the sky coordinates of the \emph{propagation}
direction; the GW source has antipodal coordinates
$\pi-\theta, \phi+\pi$.

Note that for any GW propagation direction $\Omg$, one can form 
an alternative orthonormal set of unit vectors 
by rotating $\hat m$ and $\hat n$ through an angle $\psi$ (called the 
{\it polarization angle}) in
the plane orthogonal to $\Omg$, as shown in the bottom right panel
of Figure~\ref{f:plane_wave_polarizations}.  We leave it as an
exercise for the reader to show that such a rotation of $\hat m$ and
$\hat n$ leads to a new set of polarization tensors
\be
\left[
\begin{array}{c}
e'{}^{+}_{ij}
\\
e'{}^{\times}_{ij}
\end{array}
\right]
=
\left[
\begin{array}{cc}
\phantom{-}\cos 2\psi & \sin 2\psi
\\
-\sin 2\psi & \cos 2\psi
\end{array}
\right]
\left[
\begin{array}{c}
e^+_{ij}
\\
e^\times_{ij}
\end{array}
\right]
\,,
\label{e:rot_pol_real}
\ee
or, equivalently,%
\footnote{As illustrated by comparing \eref{e:rot_pol_real} and \eref{e:rot_pol_comp},
it is often more convenient to work with complex quantities and then
take their real and imaginary parts, as opposed to working
with just the real polarizations.
This usually leads to equations that have half as many terms.}
\be
e'_{ij}(\Omg,\psi) = \exp(-i2\psi)\, e_{ij}(\Omg)\,.
\label{e:rot_pol_comp}
\ee
This change of polarization basis is a spin-two gauge
transformation that leaves the metric perturbations $h_{ij}(t,\vec x)$ 
unchanged.  This is because the 
complex amplitude is multiplied by a phase
which cancels that of \eref{e:rot_pol_comp}:
\be h'(u) =
  \exp(-i2\psi) h(u),
\label{e:rot_strain_comp}
\ee
as can be seen from \eref{e:h_ij_comp}.
Note that this mixes $h^+(u)$ and $h^\times(u)$:
$h'{}^{+}$ and $h'{}^{\times}$ are linear combinations of
 $h^+$ and $h^\times$.
The gauge transformation is local, because if there are many sources
with different $\Omg$, then $\psi$ may be picked differently for each
of these.  The spin-two interpretation comes from quantum mechanics,
where it indicates that the phase of the wave function rotates at
twice the rate of the physical coordinate axes.

A particularly interesting case is the GWs emitted by a
pair of SMBHs which are in a circular orbit with constant angular
frequency, with an orbit that lies in the plane of the sky. This means
that $\Omg$ is parallel to the orbital angular momentum.  The
resulting GWs have components
\be
h^+(t) = A\cos(2\pi f_0 t+\phi_0)\,,
\quad
h^\times(t) = A\sin(2\pi f_0 t+\phi_0)\,,
\label{e:hA_circ}
\ee
where $A$ is a constant which is inversely proportional to the distance from Earth or pulsars to the SMBHs.
The complex waveform is
\be
h(t)\equiv h^+(t)+ih^\times(t)
= A \exp\left(i(2\pi f_0 t+\phi_0)\right)\,.
\label{e:h_circ}
\ee
Using \eref{e:h_ij_comp} and \eref{e:rot_pol_comp}, it follows that%
\footnote{As written in \eref{e:h_ij_circ}, 
the time dependence of the metric perturbations
$h_{ij}(t,\vec x)$ has been moved from the functions 
$h^A(t)$ into the polarization tensor 
$e'{}^+_{ij}(\Omg, \psi(t))$, via the time dependence of $\psi(t)$.
Normally, the polarization
tensors are taken to be time independent.}
\be
h_{ij}(t,\vec 0) = A\, e'{}^+_{ij}(\Omg, \psi(t))\,,
\quad
\psi(t) \equiv \frac{1}{2}\left(2\pi f_0 t+\phi_0\right)\,,
\label{e:h_ij_circ}
\ee
which describes a {\it circularly polarized} GW.
The unit vectors $\hat m'$ and $\hat n'$ 
defining $e'{}^+_{ij}(\Omg, \psi(t))$ rotate uniformly in 
the plane orthogonal to $\Omg$ with angular velocity 
$\d\psi/\d t = \pi f_0$,
which is the orbital angular velocity of the SMBH pair;
the GW angular frequency is twice this.

For a circularly polarized GW source, the two linear
polarization strain components are uncorrelated in time and have equal
time-averaged squared values:
\be
\overline{h^+(t)^2}=
\overline{h^\times(t)^2}\,,
\quad
\overline{h^+(t)h^\times(t)}=0\,.
\label{e:unpolarized}
\ee
Here, overbar denotes an average 
\be
\overline{F(t)} \equiv \frac{1}{T}\int_{-T/2}^{T/2} \d t\> F(t)
\label{e:time-avg}
\ee
over the observation time $T$, and
we have assumed that $T$ is commensurate with the
period of the wave, so that $f_0 T = n$, where $n \ne 0$ is an integer.

Any gravitational-wave source whose $+$ and $\times$ components
satisfy \eref{e:unpolarized} is said to be {\it unpolarized}.
At first, this seems to be a strange use of language, since
it means that a circularly polarized source is ``unpolarized''. A
better way to think about it is this: an unpolarized source has no
preferred polarization direction, as defined by the polarization
angle $\psi$, in the plane perpendicular to $\Omg$.  So, according
to this definition, a circularly polarized source {\it is}
unpolarized, whereas a linearly polarized or elliptically polarized
source is \emph{not} unpolarized.  The Hellings and Downs
curve that we will discuss later is appropriate for a GWB made
up of a superposition of unpolarized GW sources, or for a collection
of polarized sources, provided that the collection does not define
any preferred polarization angle.

\subsection{Redshift response for a single Earth-pulsar baseline}
\label{s:redshift_response}

As the GW passes the Earth-pulsar baseline, 
it affects the pulse arrival times.
The change $\Delta T$ in the arrival time of the pulse as measured at Earth
at time $t$ is obtained by projecting the metric perturbations
along the direction to the pulsar $\hat p$, and then
integrating that projection along the spacetime path of the 
pulse from emission at the pulsar to reception at Earth:%
\bnp
\bea
&\Delta T(t)=\frac{1}{2}\hat p^i\hat p^j\int_0^L\d s\> h_{ij}(t(s),\vec x(s))
\quad{\rm where}
\label{e:DeltaT(t)}
\\
&t(s) = t-(L-s)\,,\quad 
\vec x(s) = (L-s)\hat p\,.
\label{e:pulse_path}
\eea
\enp
Thus, the pulse ``samples'' the metric perturbations along its (null) path through spacetime.
Note that we do not need to
include any corrections to the straight-line path for the photon
given above, as the metric perturbations are already first-order small
and we can ignore all second- and higher-order terms in
the calculation~\cite{Sachs-Wolfe:1967}.

It turns out to be somewhat simpler mathematically to work instead
with the Doppler shift in the pulse frequency induced by the GW.
This Doppler or ``redshift" response, which we will denote by $Z(t)$, is 
obtained by differentiating the timing residual response $\Delta T(t)$ with 
respect to time:
\be
Z(t) 
\equiv \frac{\d\Delta T(t)}{\d t}
=\frac{1}{2}\hat p^i\hat p^j\int_0^L\d s\> 
\frac{\partial h_{ij}}{\partial t}(t(s),\vec x(s))\,.
\label{e:Z_def}
\ee
A redshift means an apparent lowering of the clock frequency, 
corresponding to an increase in the arrival period of the pulse,
while a blueshift is the opposite.
While ``redshift" and ``blueshift" are normally used in the context 
of visible light, they also apply to
radio signals and even to the time delays between pulses from a pulsar,
since relativistic effects are felt by all types of
clocks and oscillators.  
(A detailed derivation of this equation in the context of 
temperature fluctuations in the cosmic microwave background 
was given by Sachs and Wolfe in 
\cite{Sachs-Wolfe:1967};
see Eq.~(39) in that paper.
This was the first such occurrence of this equation that we
are aware of.)

To perform the integration and differentiation that appear in
\eref{e:Z_def}, we first bring the 
time derivative inside the integral, differentiating the complex 
function $h(u(t,s))$ that enters the expression for 
$h_{ij}(t(s),\vec x(s))$, see \eref{e:h_ij_comp} and \eref{e:pulse_path}.
Since
\be
u(t,s)\equiv t(s) - \Omg\cdot\vec x(s)
= t-(L-s)(1+\Omg\cdot \hat p)\,,
\ee
it follows that
\bea
\frac{\partial h}{\partial t} = \frac{\d h}{\d u}\frac{\partial u}{\partial t}
=\frac{\d h}{\d u}\,,
\quad
\frac{\partial h}{\partial s} = \frac{\d h}{\d u}\frac{\partial u}{\partial s}
=(1+\Omg\cdot\hat p)\frac{\d h}{\d u}\,,
\eea
leading to 
\be
\frac{\partial h}{\partial t} 
= \frac{\d h}{\d u} 
= \frac{\partial h/\partial s}{\partial u/\partial s} 
= \frac{1}{(1+\Omg\cdot \hat p)}\frac{\partial h}{\partial s}\,.
\ee
The
expression
\eref{e:Z_def}
can now be trivially integrated over $s$.
Since the prefactor $1/(1+\Omg\cdot \hat p)$ does not depend on $s$,
the integral is just the difference of 
$h(u(t,s))$
at the endpoints:
\be
Z(t)
= \frac{1}{2} \frac{{\hat p}^i  {\hat p}^j }{1+{\Omg} \cdot {\hat p}}\Bigl[ h_{ij}\bigl(t, \vec 0
 \bigr) - h_{ij}\bigl( t - L, L \hat p \bigr) \Bigr]\,.
\label{e:Z_earth_pulsar}
\ee
This is the difference of the metric perturbations at the
Earth at the instant when the pulse was received [$s=L \implies$
spacetime point $(t ,\vec 0)$] and at the pulsar at the instant
when the pulse was emitted [$s=0 \implies $ spacetime point $(t-L ,L\hat p)$], 
projected onto the pulsar direction.  The two terms
are called the {\it Earth term} and {\it pulsar term}, respectively,
and are explicitly shown in \Fref{f:one_pulsar_geom}.

This difference between the Earth and pulsar terms has a simple
physical interpretation.  As explained in \cite{Narlikar:1994,
  Bunn-Hogg:2009, Romano-Creighton:2024}, the redshift \eref{e:Z_earth_pulsar} can be
interpreted as a Doppler shift resulting from a relative velocity
between the Earth and the pulsar, projected along the line of sight to
the pulsar.  This projected relative velocity is obtained by
parallel-transporting the pulsar 4-velocity vector along the null
geodesic traversed by the radio pulse.  Via the Doppler effect, this
nonzero relative velocity redshifts or blueshifts the photon frequency.
If both source and receiver have the same projected velocity, then the
Earth and pulsar term cancel, and the GW has no net effect on the
photon redshift/pulsar timing residual.

We can write the above expression for the redshift
response $Z(t)$ as 
\be
Z(t)=  \sum_{A}\Delta h^A(t,L\hat p) F^A (\Omg)
=\Re\left\{\Delta h(t,L\hat p)F^* (\Omg)\right\}\,,
\label{e:Z(t)}
\ee
where $\Delta h^A$ are the the Earth-to-pulsar differences in retarded waveforms:
\be
\label{e:Deltah^A}
\Delta h^A(t,L\hat p)\equiv h^A(t) - h^A(t-L(1+\Omg\cdot \hat p))\,,
\ee
and the $F^A$ are geometrical projection factors
\be
\label{e:F^A}
F^A(\Omg) \equiv \frac{1}{2} \frac{{\hat p}^i  {\hat p}^j }{1+{\Omg} \cdot {\hat p}}e_{ij}^A(\Omg)\,,
\quad A=+,\times\,.
\ee
The corresponding complex quantities $\Delta h(t,L\hat p)$ and $F(\Omg)$ are defined by
$\Delta h\equiv\Delta h^+ + i \Delta h^\times$ and $F \equiv F^+ + i F^\times$.

At first glance it appears that the geometrical projection factor
$F^A(\Omg)$ diverges if the GW source is located behind
  the pulsar ($\Omg=-\hat p$), since the denominator $1+\Omg\cdot\hat
p$ in~\eref{e:F^A} then vanishes.  But the reader can easily verify
that this zero in the denominator is canceled by a zero from the
projection $\hat p^i \hat p^j e^A_{ij}(\Omg)$ when $\hat p$ is
proportional to $\Omg$.  In this $\Omg \to -\hat p$ case,
  the response $|F| \to 1$ is finite, as can be seen clearly from
  \eref{e:redshiftCircPolSource5}. Nevertheless, the
  redshift $Z(t)$ vanishes: because $1+\Omg\cdot\hat p =0$, the two
  functions in \eref{e:Deltah^A} are both evaluated at $t$, and
  cancel. Physically, the radio pulse ``surfs'' on the GW, so the
  Earth and pulsar term are identical and their difference is zero.
  However, if the GW direction $\Omg$ is displaced slightly away from
  $-\hat p$, the response has its maximum value as shown by the nearly
  vertical ``lobes" in the rightmost panel of \Fref{f:G_F}.

\subsection{Transfer functions and geometric projection factors}
\label{s:TransferGeometricAntenna}

To visualize the response of a pulsar to a GW source, consider the
simplest case: a circularly polarized GW source at sky position $-
\Omg$ radiating at a fixed GW frequency $f$. In a linear polarization
basis the waveform is given by \eref{e:hA_circ}, and in a circular
($\equiv$ complex) polarization basis it is given by \eref{e:h_circ}.
Without loss of generality, the amplitude $A \ge 0$.  From
\eref{e:Z(t)}, the redshift at Earth for a pulsar in direction $\hat
p$ is
\begin{equation}
\label{e:redshiftCircPolSource1}
Z(t) =  \Re \Bigl( A  {\rm e}^{i(2\pi f t+\phi_0)}
   \Bigl[ 1 - {\rm e}^{-i2\pi f L(1+\Omg\cdot\hat p)} \Bigr]
   \Bigl[ F^+(\Omg) - i F^\times(\Omg)  \Bigr] \Bigr) \, .
\end{equation}
The first quantity in square brackets is the \emph{redshift transfer
function}, which is the difference between the Earth term (unity) and
the pulsar term (complex phase):
\be
{\cal T}(\alpha, \Omg\cdot \hat p)\equiv 
1 - e^{-i2\pi \alpha(1+\Omg\cdot\hat p)}\,  .
\label{e:redshift_transfer}
\ee
The dimensionless parameter $\alpha \equiv fL = L/\lambda$ we have
introduced is the ratio of the arm-length (i.e., the distance from
Earth to pulsar) to the wavelength of the GW.

Return to the redshift in \eref{e:redshiftCircPolSource1}.  Let $r \ge
0$ denote a real modulus and $\phi \in [0, 2\pi)$ denote a real phase,
  so the redshift is
\begin{equation}
\label{e:redshiftCircPolSource2}
Z(t) =  \Re \Bigl( r  {\rm e}^{i(2\pi f t + \phi)}  \Bigr) = r \cos (2\pi f t + \phi) \, .
\end{equation}
The squared modulus can be read directly from
\eref{e:redshiftCircPolSource1}, and is
\begin{equation}
\label{e:redshiftCircPolSource3}
r^2  =  A^2 \left|  1 - {\rm e}^{-i2\pi f L(1+\Omg\cdot\hat p)}  \right|^2
\ |F(\Omg)|^2
\end{equation}
Thus, the redshift oscillates sinusoidally with time according to
\eref{e:redshiftCircPolSource2}, with an overall amplitude
\begin{equation}
\label{e:redshiftCircPolSource4}
r  =  A \bigl|
{\cal T}(\alpha, \Omg\cdot \hat p)
\bigr| |F(\hat\Omega)|,
\end{equation}
where
\begin{equation}
  \label{e:redshiftCircPolSource5}
|F(\hat\Omega)| \equiv \sqrt{ F^+(\Omg)^2  + F^\times(\Omg)^2 } 
= \frac{1}{2}(1-\Omg\cdot\hat p)
\end{equation}
is the quadrature combination of the pulsar's responses to the two
polarizations.
We recommend that the reader verify the last equality in 
\eref{e:redshiftCircPolSource5};
see also \cite[Eq.~(2.7)]{allen2024angular}.

Is the redshift response for a source which is unpolarized [as defined
  by \eref{e:unpolarized}] described by the same quadrature
combination?  No, not unless the definition of ``unpolarized'' is
generalized.  This can be seen by computing the average squared
redshift $\overline{Z^2(t)}$, 
where the overbar denotes the time average in \eref{e:time-avg}.
This is proportional to  $F^+(\Omg)^2 + F^\times(\Omg)^2$
if
\be
\overline{h^+(t) h^+(t+T)}= \overline{h^\times(t) h^\times(t+T)}\,,
\text{ and } 
\overline{h^+(t) h^\times(t+T)}=0\,,
\label{e:unpolarizedGen}
\ee 
for an arbitrary value of $T$.  If $h^+$ and $h^\times$ are
ergodic processes, then these auto-correlation functions are related
to the spectrum and cross spectrum of the two polarizations: the
mean-squared redshift response will be described by the quadrature
combination if $h^+$ and $h^\times$ have the same power spectrum and a
vanishing cross spectrum.

The quadrature combination is also relevant for polarized sources, if
there are no preferred orientations.  To see this, first consider a
\emph{single} polarized source, defined by some complex waveform $h(u)
= h^+(u) + i h^\times (u)$. From \eref{e:Z(t)}, the redshift at Earth
is $Z(t) = \Re [ \Delta h(t, L \hat p) F^*(\Omg) ]$.  How would that
be changed if the axes which define the source's polarization are
rotated through an angle $\psi$ as shown in the bottom right panel of
Figure~\ref{f:plane_wave_polarizations}?  Because this does not change
$\Omg$ or the direction $\hat p$ to the pulsar, it only affects the
redshift via $F$, which is modified because the complex polarization
tensor appearing in $F$ is transformed according to
\eref{e:rot_pol_comp}.  Let $C_R$ and $C_I$ denote the real and
imaginary parts of the complex quantity $ C = C_R + i C_I = \Delta
h(t, L \hat p) F^*(\Omg)$.  Then the redshift for this new source,
with rotated polarization axes, is
\begin{equation}
  \label{e:redshiftUnpolSource2}
  Z(t) = \Re \left[ C {\rm e}^{i 2 \psi} \right] = C_R \cos( 2 \psi) - C_I \sin( 2 \psi).
\end{equation}
The average squared value of this quantity (uniformly averaged over all polarization orientations) is
\begin{equation}
  \label{e:redshiftUnpolSource2.5}
  Z^2_{\rm rms}(t) \equiv  \frac{1}{2 \pi} \int_0^{2 \pi} \!\! \!\!\! \d \psi \, Z^2(t) = \frac{1}{2} \left( C_R^2 + C_I^2 \right) =
  \frac{1}{2} \left| C \right|^2.
\end{equation}
Taking the square root of both sides, we find that the rms redshift is
\begin{equation}
    \label{e:redshiftUnpolSource3}
 Z_{\rm rms} =  \frac{1}{\sqrt{2}}  \bigl| h(t) - h(t-L(1 + \Omg\cdot\hat p)) \bigr| \bigl|F(\Omega) \bigr|.
  \end{equation}
Thus, these quadrature combinations characterize the response of a PTA
to an unpolarized source, or to a polarized source whose orientation
is random.

\begin{figure}[htbp!]
\begin{center}
\includegraphics[width=0.5\textwidth]{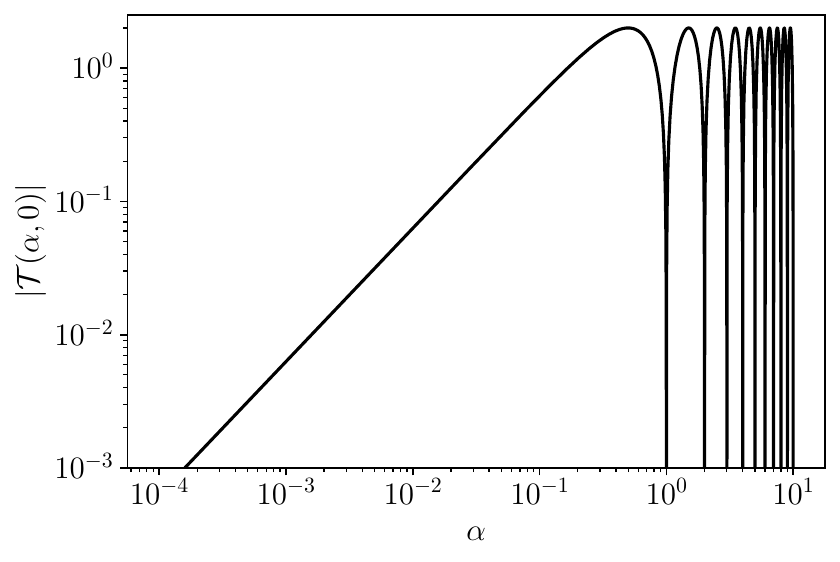}
\caption{Modulus of the redshift transfer function
${\cal T}(\alpha, \Omg\cdot \hat p)$ for a normally incident
($\Omg\cdot\hat p=0$) GW.  
It vanishes if the arm length is a multiple of the GW wavelength, $\alpha\equiv fL = n$ for $n=1,2,3,\dots$.}
\label{f:redshift_transfer}
\end{center}
\end{figure}

The redshift \eref{e:Z(t)} is linear in the GW strain, and any GW
strain can be decomposed into a sum of (left and right)
circularly polarized plane waves of constant frequency.  Hence, we can
understand the response of a PTA to any GW signal by understanding the
response of the PTA to a circularly polarized GW of constant
frequency. In light of \eref{e:redshiftCircPolSource1} it is
convenient to define response functions
\be
R^A(f,{\Omg}) 
\equiv \left[1 - e^{-i 2\pi fL(1+\Omg\cdot \hat p)}\right] F^A(\Omg)
=  {\cal T}(\alpha, \Omg\cdot \hat p) F^A(\Omg) \, .
\label{e:R^A}
\ee
%
The redshift transfer function ${\cal T}$ vanishes if the pulse from
the pulsar traverses an integer number of cycles of GW-induced
stretching and squeezing, because these integrate to give zero net
timing delay.  At fixed GW frequency $f$, this occurs for ${\rm
  int}[2\alpha]$ different incident directions, given by values of
$\Omg\cdot \hat p$ that satisfy
\be
1+\Omg\cdot \hat p = n/\alpha
\quad{\rm where}\quad
n=1,2,\dots, {\rm int}[2\alpha]\,.
\label{e:theta_zeros}
\ee
The limit on $n$ arises because the left-hand side of~\eref{e:theta_zeros} lies
in the interval $[0,2]$.  However, for a given GW direction (fixed
$\Omg\cdot\hat p$) there are an infinite number of $\alpha$ values 
(or frequencies) at which $\cal T$ vanishes.  For normal incidence
of the GW (i.e., $\Omg\cdot\hat p=0$), 
these zeroes lie at $\alpha=n $ for $n=1, 2, \dots $, as shown in
\Fref{f:redshift_transfer}.

For a short-arm LIGO-like detector, the redshift transfer
function~\eref{e:redshift_transfer} and the response
function~\eref{e:R^A} take simple forms.  Expanding the
exponential in \eref{e:redshift_transfer} to first order in $\alpha\ll
1$ gives ${\cal T} \simeq i2\pi\alpha(1+\Omg\cdot\hat p)$, which
implies (using $\alpha\equiv fL$)
\be
R^A(f,\Omg) 
\simeq i 2\pi fL\, G^A(\Omg)\,,\quad
G^A(\Omg) \equiv \frac{1}{2}
\hat p^i \hat p^j e_{ij}^A(\hat\Omega)\,.
\label{e:G^A}
\ee
The geometrical projection factors $G^A(\Omg)$, where $A=+,\times$, 
are just the {\it strain} (i.e., $\Delta L/L$) 
response functions for a short 
one-arm, one-way LIGO-like detector.
Note that this short-arm response function is symmetric
under the interchange $\hat p\rightarrow-\hat p$, unlike
$F^A(\Omg)$.
This implies that if the distance from the Earth to a pulsar 
were much less than the GW wavelength, then the timing response of
that Earth-pulsar baseline $\hat p$ would be exactly the 
same as that for a pulsar in the opposite sky direction, 
$-\hat p$.
This symmetry can be seen graphically in the leftmost panels of 
Figures~\ref{f:antenna_patterns} and \ref{f:G_F}.

\subsection{Visualizing the redshift response (antenna pattern functions)}
\label{s:antenna_patterns}

It is helpful to visualize the response of a single Earth-pulsar
baseline to an ``unpolarized source'' in the sense described above.
From \eref{e:redshiftCircPolSource4}, for a unit-amplitude source
($A=1$) the relevant quantity is the {\it polarization-averaged}
response function
\be
{\cal R}(f,\Omg)
=\sqrt{|R^+(f,\Omg)|^2 + |R^\times(f,\Omg)|^2}
= | {\cal T}(\alpha, \Omg\cdot\hat p)| \, |F(\Omg)|\, ,
\label{e:R_avg}
\ee 
with $\alpha\equiv fL$ as previously defined, and $|F(\Omg)|$ given by
\eref{e:redshiftCircPolSource5}. To examine the short-arm
limit, we also define the {\it quadrature combination}
\be
|G(\Omg)|\equiv \sqrt{G^+(\Omg)^2 + G^\times(\Omg)^2}
\label{e:G_avg}
\ee
for the response in the short-arm limit, ignoring the 
prefactor $i2\pi fL$ relating $R^A(f,\Omg)$ and $G^A(\Omg)$,
see \eref{e:G^A}.

Without loss of generality, consider a pulsar located in the $\hat p =
\hat z$ direction
\be
|F(\Omg)| = \frac{1}{2}(1+\cos\theta)\,,
\quad
|G(\Omg)| = \frac{1}{2}\sin^2\theta\,,
\qquad
\label{e:G_F_explicit}
\ee
where $-\Omg\cdot\hat p = \cos\theta$ is the cosine of the
angle of the GW source relative to the pulsar direction.
You will need to
use~\eref{e:e+}, \eref{e:ex}, \eref{e:Omg}, \eref{e:m}, \eref{e:n}, 
\eref{e:G^A}, and \eref{e:F^A}
to obtain these explicit expressions.  As discussed above, 
$|F(\Omg)|$ is finite as $\theta \to 0$ and vanishes as $\theta \to \pi$,
while $|G(\Omg)|$ vanishes at both $\theta=0,\pi$
(see the middle and left panels of \Fref{f:G_F}).

\begin{figure}[htbp]
\begin{center}
\includegraphics[width=0.9\textwidth]{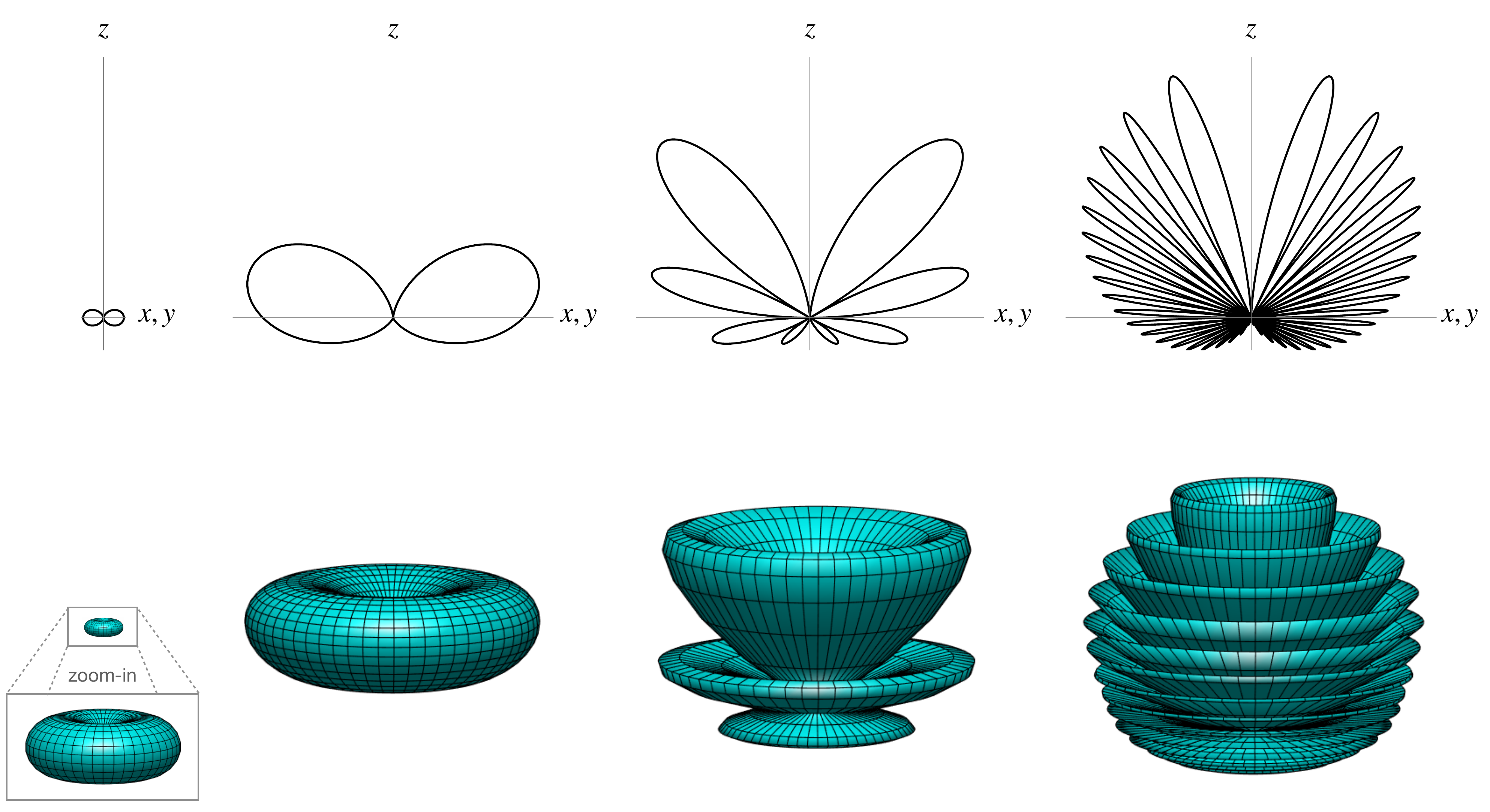}
\caption{ Antenna pattern functions ${\cal R}(f,\Omg)$ for the polarization-averaged
  redshift response of an Earth-pulsar baseline having $\hat p=\hat
  z$, evaluated for different fixed values of $\alpha\equiv fL$ (from
  left to right, $\alpha=0.05$, 0.5, 2, 10).  The top row of plots are
  cross-sections: the antenna pattern functions are axially symmetric
  about the vertical $z$-axis. For example, the leftmost pattern
  resembles a doughnut with no hole, as shown in the bottom row of plots,
  which are the corresponding antenna patterns viewed from slightly
  above the $xy$ plane. }
\label{f:antenna_patterns}
\end{center}
\end{figure}

\begin{figure}[htbp]
\begin{center}
\includegraphics[width=0.9\textwidth]{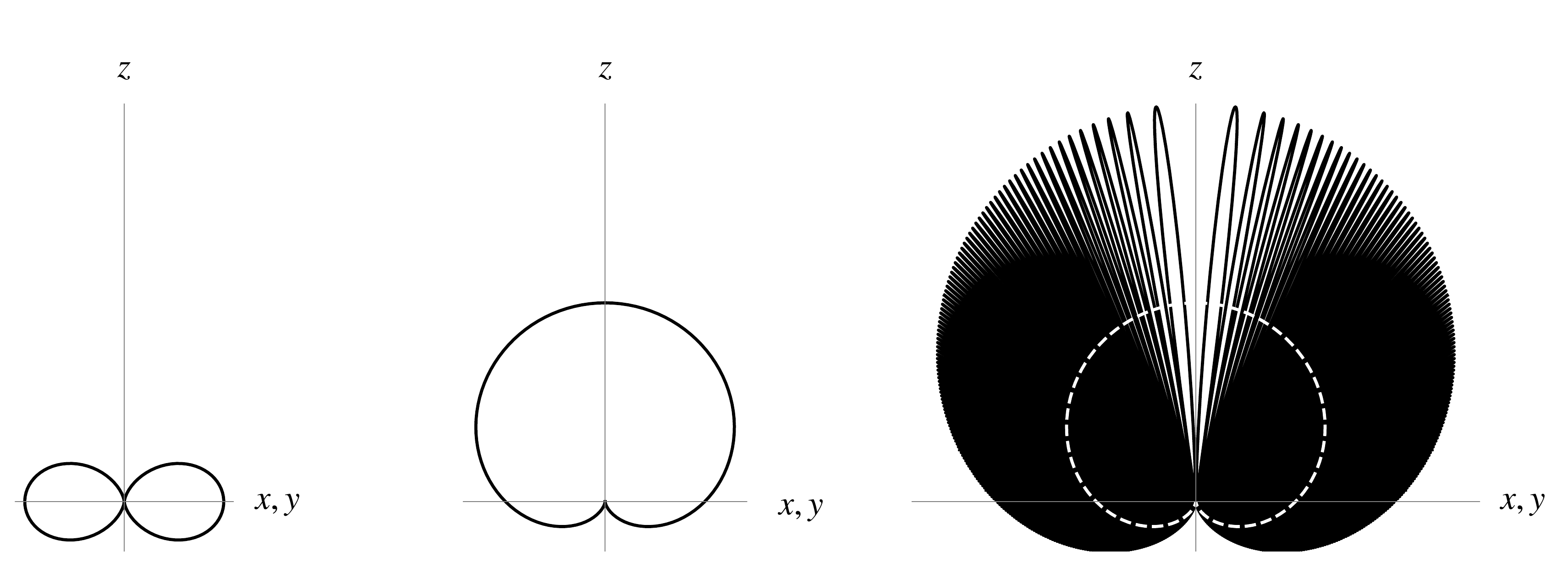}
\caption{Antenna pattern functions for the polarization-averaged
geometrical projection factors $|G(\Omg)|$ (left panel)
and $|F(\Omg)|$ (middle panel).
The right panel compares $|F(\Omg)|$ (white dashed
curve) with the full polarization-averaged timing 
response function ${\cal R}(f,\Omg)$ for $fL=100$.
All of these antenna pattern functions are axially 
symmetric around the vertical $z$-axis.
}
\label{f:G_F}
\end{center}
\end{figure}

For fixed frequency $f$ and fixed distance $L$ to a pulsar, the {\it
shape} of the polarization-averaged response function ${\cal R}(f,\Omg)$
depends only on the relative orientation of the GW propagation
direction $\Omg$ and the direction to the pulsar $\hat p$.
The shape of ${\cal R}(f,\Omg)$ is often called an {\it antenna pattern},
because the magnitude of ${\cal R}$ is the ``strength" of the response.
\Fref{f:antenna_patterns} shows plots of the antenna patterns
${\cal R}(f,\Omg)$ for a pulsar located in the $\hat z$ direction for
different fixed values of $\alpha\equiv fL$.  The left panel shows the antenna
pattern in the ``short-arm" LIGO-like limit, $\alpha\ll 1$, which is
proportional to $\sin^2\theta$, as shown in
\eref{e:G^A} and~\eref{e:G_F_explicit}. Note that the number of
lobes of the antenna pattern increases as $\alpha$ increases, and that
the overall magnitude of the antenna pattern function is proportional
to $\alpha$ for $\alpha\ll 1$. 
Both of these behaviors are a consequence of the form
of the redshift transfer function given in \eref{e:redshift_transfer}
and seen in \Fref{f:redshift_transfer}.

Recall that for most pulsars, $\alpha$ will be of order 100 or 
more, since the distance to typical pulsars is of order a kpc 
(so a few thousand light-years) or more, while the 
GW wavelengths are only of order tens of light-years.
The rightmost panel of \Fref{f:G_F} is a plot of ${\cal R}(f,\Omg)$ 
appropriate for the (semirealistic) case $\alpha=100$.
Also plotted in this figure are the polarization-averaged 
antenna pattern functions for the geometrical projection 
factors $|G(\Omg)|$ (left panel) and $|F(\Omg)|$ (middle panel).
Note that $|F(\Omg)|$ is half of the ``envelope" 
of the full response function ${\cal R}(f,\Omg)$, illustrated 
by the dashed white curve in the rightmost panel plot.

\begin{figure}[htbp]
\begin{center}
\includegraphics[width=0.6\textwidth]{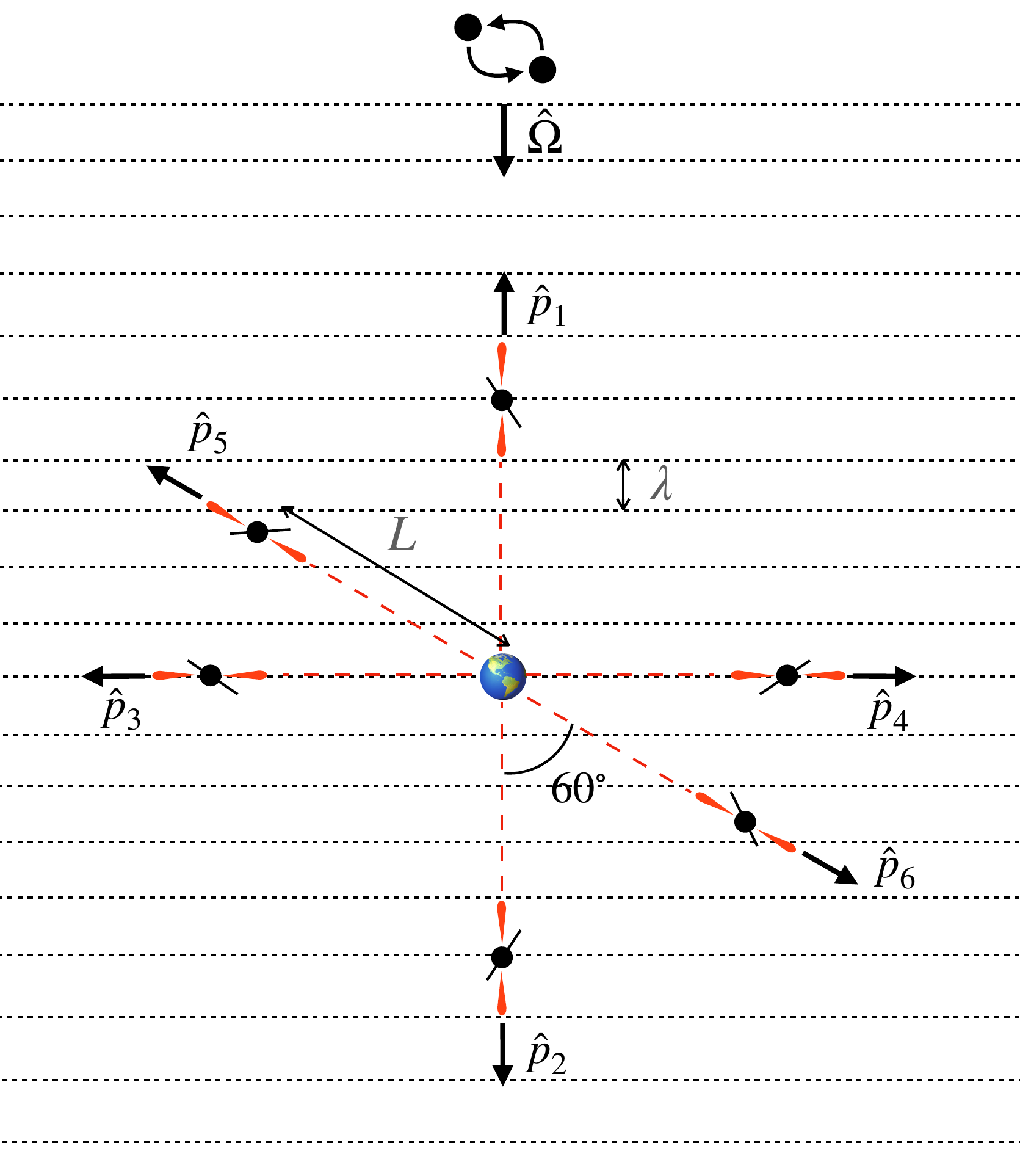}
\caption{Six hypothetical pulsar locations relative to the 
Earth and a monochromatic GW propagating in direction $\Omg$.
All the pulsars lie in a plane, and are at distances 
$L=5\lambda$ with respect to the Earth, where $\lambda$ is 
the wavelength of the GW.}
\label{f:array}
\end{center}
\end{figure}

These differences in antenna patterns and responses for
  ``short-arm" and ``long-arm" GW detectors have close equivalents in
  the context of electromagnetic (EM) antennas and phased arrays.
  Those which are small compared to the EM wavelength are analogous to
  short-arm GW detectors, and vice versa. Because of reciprocity, it
  does not matter if the antennas are receiving or broadcasting.

  A simple example is the center-fed linear antenna.  This a straight
  wire of length $d$, laid along the $z$-axis between $-d/2 < z <
  d/2$, and broken open at the origin, where the two halves are
  driven by a voltage difference that varies sinusoidally with
  time. This creates a current distribution along the wire which is
  sinsoidal with respect to both time and space, and which vanishes at
  the ends of the wire, see \cite[Sec.~9.4]{Jackson2nd} for
  details. In the radiation zone $r\gg d$, $r\gg\lambda$, where $r$ is
  the distance from the antenna, the electric field $\vec E$ has an
  azimuthally-symmetric magnitude proportional to
  \be
  \label{e:EMantenna}
| \vec E| \propto \frac{\cos\left(\frac{\pi d}{\lambda}\cos\theta\right) - 
\cos\left(\frac{\pi d}{\lambda}\right)}{\sin\theta} \frac{1}{r}\,.
\ee
If the antenna is long, $d \gg \lambda$, then this radiation pattern
has multiple lobes, similar to that shown in the rightmost panel of
Figure~\ref{f:G_F} for a PTA.  If the antenna is short, $d\ll
\lambda$, then \eref{e:EMantenna} approaches the single-lobed
radiation pattern of a short EM dipole antenna $r |\vec E| \propto
(\pi d/\lambda)^2\,\sin\theta$.  This is similar to the response shown
in the left-hand panel of Figure~\ref{f:G_F} for a short-arm GW
detector.

In the EM case, the multiple lobes are due to constructive and
destructive interference from the sinusoidally-varying current
distribution along the length of the long antenna.  In the GW PTA
case, the multiple lobes are due to constructive and destructive
interference of the sinusoidally-varying GW amplitude along the photon
path.

To end this section, we leave it as an exercise for the 
reader to calculate the values of the various antenna 
pattern functions $|G(\Omg)|$, $|F(\Omg)|$, and ${\cal R}(f,\Omg)$ 
for the different pulsar locations and GW shown in \Fref{f:array}.
As in \Fref{f:one_pulsar_geom}, there is a single GW point
source in direction $-\Omg$ relative to Earth, sufficiently
far away that the wavefronts are planar when they reach the 
vicinity of the Earth and the pulsars.
We also assume that the GW is monochromatic with
$fL \equiv L/\lambda=5$, where $\lambda$ is the GW wavelength 
and $L$ is the distance of the pulsars from Earth.
The values of the various antenna patterns are given in \Tref{t:array}.
\begin{table}
\caption{\label{t:array}Values of the different geometrical and
  antenna pattern functions for the GW source and pulsar locations
  shown in \Fref{f:array}.}
\begin{indented}
\item[]\begin{tabular}{@{}lcccccc}
\br
Pulsar label & $a=1$ & $a=2$ & $a=3$ & $a=4$ & $a=5$ & $a=6$ \\
Pulsar direction & $\hat p_1$ & $\hat p_2$  & $\hat p_3$  & $\hat p_4$  & $\hat p_5$  & $\hat p_6$ \\
$|G_a(\Omg)|$ & 0 & 0 & 1/2 & 1/2 & 3/8 & 3/8\\
$|F_a(\Omg)|$ & 1 & 0 & 1/2 & 1/2 & 3/4 & 1/4\\
${\cal R}_a(f,\Omg)$ & 0 & 0 & 0 & 0 & 3/2 & 1/2\\
\br
\end{tabular}
\end{indented}
\end{table}

\subsection{The plane wave expansion}
\label{ss:planewaveexpension}

The plane wave expansion is widely used in the literature and provides
a simple and precise way to write the pulsar response in the most
general case of interest.  If the PTA is far from the GW sources, then
the GW metric perturbations in the PTA's neighborhood may be written as a sum
of plane waves
\be
\label{e:PlaneWaveExpansion}
h_{ij}(t,\vec x) = \sum_A \int{\rm d}f \int {\rm d}\Omg \, h_A(f,\Omg)
  e^A_{ij}(\Omg)  {\rm e}^{i2\pi f(t - \Omg \cdot \vec x)}\,,
\ee
where $\int {\rm d}\Omg\equiv \int_0^{2\pi} {\rm d}\phi \int_0^{\pi} \sin\theta\, {\rm d}\theta$.
Here
and elsewhere, frequency and time integrals are over the range
$(-\infty,\infty)$, unless explicit limits are given.

One can see by inspection that $h_{ij}(t,\vec x)$ satisfies the wave equation
$(-\partial_t^2 + \nabla^2)h_{ij} = 0$ and that it is transverse
$\nabla_i h_{ij} = 0$, traceless, and synchronous.
The functions $h_A(f, \Omg) = h_A^*(-f, \Omg)$ are the complex Fourier
amplitudes of the GW sources in direction $-\Omg$ at frequency $f$.
If there is a discrete set of sources, then the integral over
$\Omg$ may be replaced by a sum over those discrete sources, each of
which has its own direction vector.

Substitute this plane wave expansion \eref{e:PlaneWaveExpansion} into
the formula for the redshift \eref{e:Z_def}, differentiate with
respect to $t$, and then integrate along the null ray coming from
pulsar ``a'' as specified in \eref{e:pulse_path}. One obtains a
compact formula for the redshift $Z_a(t)$ of an arbitrary pulsar in an
arbitrary GW background:
\bea Z_a(t)
      & = &  \sum_A \int {\rm d}f \int {\rm d}\Omg \, 
      R_a^A(f,{\Omg})
     h_A(f,\Omg)  {\rm e}^{i2\pi ft} \, .
\eea
The response function \eref{e:R^A} and the geometrical projection
factor \eref{e:F^A} carry the pulsar label ``$a$'', and depend upon
the pulsar's distance $L_a$ and sky direction $\hat p_a$.  They are
given by
\be R_a^A(f,{\Omg}) \equiv \left[1 - e^{-i 2\pi
    fL_a(1+\Omg\cdot \hat p_a)}\right] F_a^A(\Omg) \,
\ \Leftrightarrow\ R_a = R_a^+ + i R_a^\times
\label{e:R^A_a}
\ee
and
\be
\label{e:F^A_a}
F_a^A(\Omg) \equiv \frac{1}{2} \frac{{\hat p_a}^i {\hat p_a}^j
}{1+{\Omg} \cdot {\hat p_a}}e_{ij}^A(\Omg)\,, \quad A=+,\times\,.  \ee
The response function elegantly incorporates the difference between
the Earth term (unity) and the pulsar term (a complex phase).
                  
\section{Pulsar-pair correlations}
\label{s:correlated_response}

Large amplitude GWs would be directly visible in the GW detector
output.  Indeed, in the early days of pulsar timing, upper limits were
set on the amplitude of GWs by examining the timing residuals from
single pulsars~\cite{Kaspi-et-al:1994}.  The smallness of those residuals meant that the GW
amplitudes could not have been too large.

However, the GWs predicted for typical background sources are too weak
to be directly visible in the data; they cannot be separated from
noise in the measurements or other physical effects.  Fortunately,
there is a way around this: look for a {\it common} GW signal by
correlating the redshift measurements for {\it two or more}
Earth-pulsar baselines for a PTA.  With sufficient integration or
averaging, one can ``dig below the noise''.  The same method
(correlating signals from pairs of LIGO-like Earth-based detectors) is
used to hunt for a stochastic GW background at audio
frequencies~\cite{allen1997stochastic,AllenRomano,Romano2017}.

\begin{figure}[htbp]
\begin{center}
\includegraphics[width=.6\textwidth]{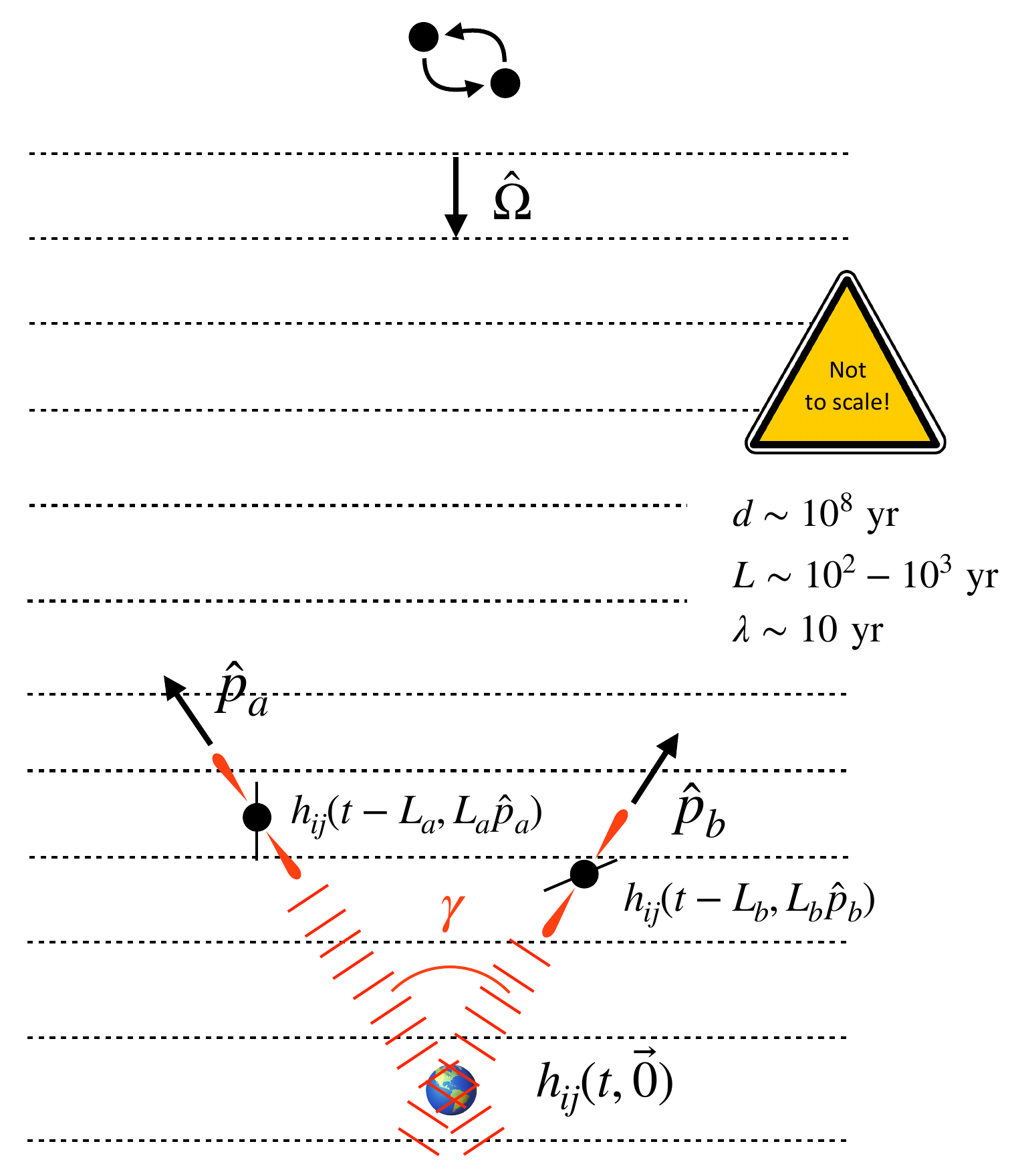}
\caption{The geometry for the pulsar-pair correlation analysis is the
  same as in \Fref{f:one_pulsar_geom}, but with two Earth-pulsar
  baselines.  The pulsars are labeled by $a$ and $b$ and are at
  distances $L_a$ and $L_b$ and directions $\hat p_a$ and $\hat p_b$
  as seen from Earth.  They are separated by angle $\gamma$ on the
  sky.  As discussed in the text, we assume that the Earth-pulsar
  distances $L_a$, $L_b$ and the distance between the two pulsars
  $|L_a\hat p_a- L_b\hat p_b|$ are much greater than the correlation
  length of the relevant GWs. Consequently, the expression for the
  expected correlation only contains the Earth-term response functions
  $F_a(\Omg)$ and $F_b(\Omg)$ for the two pulsars.}
\label{f:two_pulsar_geom}
\end{center}
\end{figure}

\subsection{Hellings and Downs correlation (source averaging)}
\label{s:source_averaging}

To analyze these correlations, we extend the calculations of
\Srefs{s:plane_wave}, \ref{s:redshift_response}, and
\ref{s:TransferGeometricAntenna} to two Earth-pulsar baselines as
shown in \Fref{f:two_pulsar_geom}.  With respect to Earth, the two
pulsars are at distances $L_a$ and $L_b$ and have sky directions $\hat
p_a$ and $\hat p_b$.

As shown in \Fref{f:two_pulsar_geom}, we begin with a single GW point
source, located very far away in direction $-\Omg$ with respect to
Earth.  The correlation between noise-free redshift measurements
$Z_a(t)$, $Z_b(t)$ for the two pulsars is given by the time-averaged
product over the observation time
\be
\rho_{ab}
\equiv\overline{Z_a(t)Z_b(t)}
\equiv \frac{1}{T}\int_{-T/2}^{T/2} \d t\> Z_a(t) Z_b(t)\,.
\label{e:rho_ab}
\ee
To study the properties of the correlation $\rho_{ab}$, particularly
for an ensemble of GW sources, it is helpful to work in the frequency
domain.

The frequency-domain representation for a single GW source is a
special case of the general plane wave
expansion~\eref{e:PlaneWaveExpansion}, where the Fourier amplitudes
vanish except for a single direction $\Omg$.  Alternatively, and
equivalently, we can return to the formulation~\eref{e:h_ij_real} for
a single source, and define Fourier amplitudes $ \tilde h^A(f)$ for
the source waveforms by
\be
\tilde h^A(f) \equiv \int \d t\> e^{-i2\pi ft} \, h^A(t)
\ \Leftrightarrow\ 
\tilde h(f) = \tilde h^+(f) + i \tilde h^\times(f) 
\,,
\label{e:htildedef}
\ee
where we give both the linear and complex polarization forms.  The redshift of
pulsar $a$ may then be written as
\bea
\label{e:Ztilde}
 Z_a(t) & =  \int \d f \> e^{i2\pi ft}\sum_A \tilde h^A(f) R_a^A(f,\Omg) \\
  \nonumber
& = \Re\left[\int \d f\> e^{i2\pi ft}\, \tilde h(f) R_a^*(f,\Omg) \right] 
.
\eea
  The
redshift of pulsar $b$ is given by the same expression as in
\eref{e:Ztilde}, with $a \rightarrow b$.

The right-hand side of \eref{e:rho_ab} contains 16 terms, because each
pulsar redshift has contributions from two polarization components
evaluated at both the Earth and pulsar, so there are four terms per
pulsar.  All of these terms contribute to $\rho_{ab}$ for a fixed set
of GW sources and a fixed pair of pulsars.

However, for calculating the {\it mean} correlation for a pair of
distinct Earth-pulsar baselines, only four terms survive, as described
in detail in, e.g., \cite{ABCPS, AllenHDVariance}.  The argument that
one uses to show this depends on the averaging procedure used to
define the mean.

The first form of averaging, employed by Hellings and Downs~\cite{HD},
keeps the pulsars fixed and averages the correlation $\rho_{ab}$ over
GW source directions $\Omg$, assuming that the sources are {\it
unpolarized} and {\it isotropic} on the sky.  In doing
this, Hellings and Downs also (implicitly) assumed that distinct
sources were uncorrelated, because they dropped cross-terms
arising from the interference between different GW sources.  This
is valid if one computes an average over many different universes,
but such terms are present in any given universe and cause
deviations from that average, as discussed in Question 7.

By performing the all-sky average for an unpolarized signal, the
Earth-pulsar and pulsar-pulsar terms enter the expected correlation
via integrals over $\Omg$ of terms proportional to the
auto-correlation function $C(\tau)$ of the GW signal%
\footnote{The auto-correlation function $C(\tau)$ is the 
Fourier transform of the power spectrum of the 
GW signal, 
$C(\tau) \equiv \int \d f\,e^{i2\pi f\tau} |\tilde h(f)|^2\big/\int \d f'\, |\tilde h(f')|^2$.}
evaluated at $\tau=\tau_a$, $\tau_b$, and $\tau_a-\tau_b$, where
$\tau_a \equiv L_a(1+\Omg\cdot\hat p_a)$ and
$\tau_b \equiv L_b(1+\Omg\cdot\hat p_b)$.
As shown in \cite{AllenHDVariance}, if
we assume that
the coherence time 
of the GW signal is short compared to $\tau_a$, $\tau_b$, and 
$|\tau_a-\tau_b|$, then these terms can be ignored relative to
the Earth-Earth terms [see (C13) and Figures~8 and 9 
in Appendix C of \cite{AllenHDVariance} for details].
However, for the correlation of a single pulsar with 
itself (i.e., $a=b$), it immediately follows that $\tau_a-\tau_b=0$.
Thus, the auto-correlation function is evaluated at 0, and hence 
doubles the contribution of the Earth-Earth term for
a stationary GW background.

The final result of this {\it source-averaging} approach,
assuming that the coherence length is small as described above, is
\be
\langle \rho_{ab}\rangle =
h^2 \muu(\g)[1+\delta_{ab}]\,,
\label{e:source_avg}
\ee
where
\bnp
\bea
h^2 \equiv \frac{1}{2}\left ( \overline{(h^+)^2  + (h^\times)^2} \right) = \frac{1}{2} \overline{|h|^2} \,,
\label{e:h^2}
\\
\muu(\gamma)
\equiv\frac{1}{4\pi}\int\d\Omg\>
\sum_A F^A_a(\Omg) F^A_b(\Omg)
=
\frac{1}{4\pi}\int\d\Omg\> F_a(\Omg) F^*_b(\Omg)
\,.
\label{e:muu_def}
\eea
\enp
Here, $h^2$ is the {\it squared GW strain} 
and $\muu(\g)$ is the Hellings and Downs curve, which is the
mean Hellings and Downs correlation 
for an unpolarized and isotropic GW background~\cite{HD}.
There is no contribution from the cross-polarization
terms since $\overline{h^+h^\times}=0$ for unpolarized
GW sources.
The $\delta_{ab}$ term in \eref{e:source_avg} is needed
to account for auto-correlations.

\begin{figure}[htbp]
\begin{center}
\includegraphics[width=.49\textwidth]{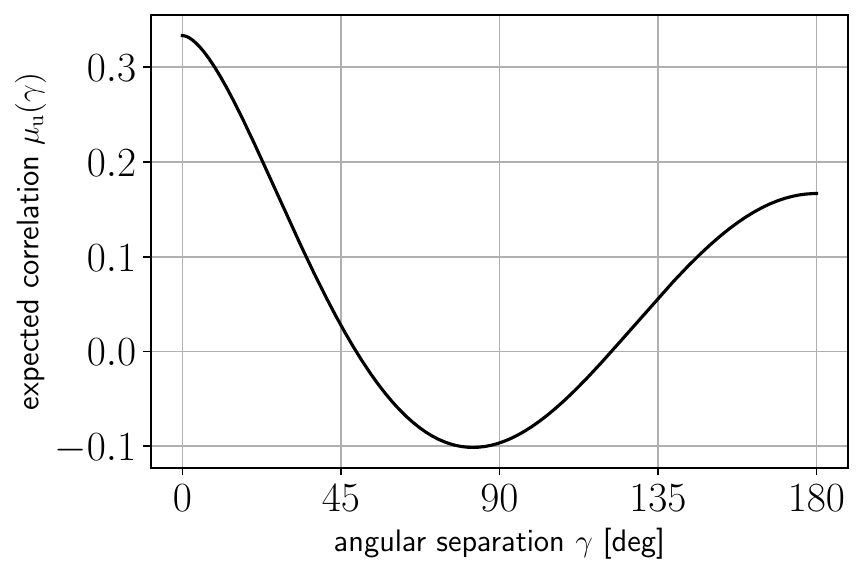}
\includegraphics[width=.49\textwidth]{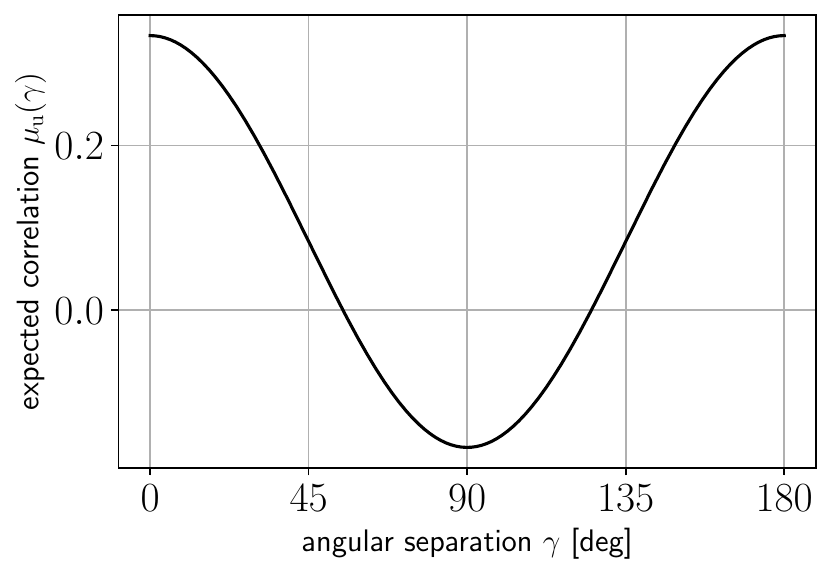}
\caption{
Left panel: Expected Hellings and Downs correlation for pairs of PTA
  pulsars.  Right panel: Expected correlation for pairs
  of short-arm LIGO-like detectors, normalized with the same 
  zero-angular-separation value of 1/3.
  Note that the zero-angular-separation value
  in the left panel corresponds to a pair of pulsars located
  close on the sky but at significantly different distances.  If they
  are much closer than a GW wavelength, or if they are the same pulsar,
  then the correlation is twice the zero-angular-separation 
  value that is shown in this plot.}
\label{f:HD_P2}
\end{center}
\end{figure}

Simple symmetry arguments show that the integral in \eref{e:muu_def}
can depend only on the angle $\g$ between the lines of sight to the
two pulsars.  So, without loss of generality, we can place pulsar $a$
on the $z$-axis and pulsar $b$ in the $xz$-plane so that $\hat
p_a=\hat z$ and $\hat p_b=\sin\g\,\hat x+\cos\g\,\hat z$.  The
polarization tensors needed to calculate $F^A_a(\Omg)$ and
$F^A_b(\Omg)$ are defined by~\eref{e:e+}, \eref{e:ex}, with $\Omg$,
$\hat m$, $\hat n$ given by \eref{e:Omg}, \eref{e:m}, \eref{e:n}.  The
integral can be evaluated analytically using contour integration, as
detailed in~\cite{ABCPS, AllenHDVariance, Jenet-Romano:2015}, giving
\be
\muu(\gamma)
=\frac{1}{3} -\frac{1}{6}\left(\frac{1-\cos\g}{2}\right)
+\left(\frac{1-\cos\g}{2}\right)
\ln\left(\frac{1-\cos\g}{2}\right)\,.
\label{e:HD}
\ee
This is an analytic expression for the Hellings and
Downs curve for pulsar timing, 
expressed as a function of the angular separation 
between the directions to pairs of pulsars. 
A plot of this expected correlation is shown in the left 
panel of \Fref{f:HD_P2}, and also as the black dashed
curve in \Fref{f:HD-NG}.

We leave it as an exercise for the reader 
to show that a similar calculation of the expected
correlation for an array of ``short-arm" 
LIGO-like detectors yields%
\footnote{We have used a proportionality sign
in \eref{e:muu_shortarm} since there are frequency-dependent
factors of $2\pi fL$ in \eref{e:G^A} that we are ignoring.}
\be
\muu(\g) \propto P_2(\cos\gamma)
=\frac{1}{2}(3\cos^2\gamma - 1)\,,
\label{e:muu_shortarm}
\ee
where $P_2(x)$ is the $l=2$ Legendre polynomial.  
For this calculation, one needs to replace the response
functions $R^A(f,\Omg)$ in~\eref{e:R^A} 
with the short-arm limiting expressions~\eref{e:G^A}.
This substitution simplifies the integrand in~\eref{e:muu_def} to
quadratics in $\sin\phi$ and $\cos\phi$ and quartics in
$\cos\theta$. 
The short-arm expected correlation is plotted in the
right panel of \Fref{f:HD_P2}.  Since the short-arm function 
involves only the $l=2$ Legendre polynomial, it is an example
of a {\it purely quadrupolar} correlation.

\subsection{Pulsar averaging}
\label{s:pulsar_averaging}

The second form of averaging, originally employed by Cornish and
Sesana~\cite{Cornish_2013}
and then further developed by the authors in \cite{AllenHDVariance,AllenRomanoOptHD},
is called {\it pulsar averaging}.
Here, for a fixed set of GW sources, one averages 
the correlation $\rho_{ab}$ over all pulsar pairs separated by 
the same angle $\gamma$, assuming that one has access to 
a large number of pulsar pairs uniformly distributed 
on the sky.
Pulsar averaging, like source averaging,
eliminates the pulsar-pulsar and Earth-pulsar terms,
provided the distances to pulsars in the same direction 
$\hat p$ are large 
compared to the GW wavelengths that PTAs are expected
to detect.
The result of pulsar averaging is a quantity
\be
\G(\g)\equiv \langle \rho_{ab}\rangle_{ab\in\g}\,,
\ee
which depends only on the angular separation $\gamma$
between pulsars $\hat p_a$ and $\hat p_b$.

Pulsar averaging corresponds to observational practice.
PTA collaborations monitor many pulsars distributed 
across the sky, so they can average together the 
correlations from all pulsar pairs lying in an angular 
separation bin centered on angle $\gamma$.
As more pulsars are added to a PTA, this 
``binned" pulsar averaging will get closer to ideal pulsar averaging,
where one has access in principle to an infinite
number of pulsar pairs uniformly distributed over the sky.
In contrast, source averaging is not observationally 
possible, since we have access to only one Universe 
and its associated (fixed) collection of GW sources.

As shown in \cite{AllenHDVariance,Cornish_2013}, pulsar 
averaging yields the same mean correlation \eref{e:source_avg} as the 
original Hellings and Downs source-averaging 
prescription for a single GW source:
\be
\G(\g) 
=h^2 \muu(\g)[1+\delta_{ab}]\,.
\label{e:G=muu}
\ee
In FAQ Q6, we demonstrate the mathematical equivalence 
of these two approaches for this simple model universe
consisting of a single GW source in a random direction.
But for multiple GW sources, the result of pulsar 
averaging is more complicated; it depends on the type 
of sources that contribute to the combined GW signal 
as described in detail in \cite{AllenHDVariance} and in 
the answer to FAQ Q7.

\section{Answers to frequently asked questions}
\label{s:answers}

Given the above background information, we are now ready to 
answer the questions that we listed in \Sref{s:intro}.

\vspace{0.1in}\hrule
\bi
\item[Q1:]
Why is the Hellings and Downs curve in \Fref{f:HD-NG} 
normalized to 1/2 at zero angular separation, but is normalized 
to $1/3$ in \Fref{f:HD_P2} and in other papers, e.g., 
\cite{HD, AllenHDVariance}? 
Does it matter?
\ei

\noindent
Answer: 
This normalization difference is an overall scale.
It is completely arbitrary and makes no difference: any
physically observable quantity (for example, the expected
time-averaged product of the timing residuals or redshift
responses of two pulsars) is
independent of this normalization.

The reason why $1/3$ and $1/2$ have appeared in the literature is
historical.  The Hellings and Downs curve was first computed 
as the {\it average} product of the redshift responses of two
pulsars to a unit-amplitude GW source, averaged over the sphere of GW
source directions as described in \Sref{s:source_averaging}. 
This average introduces a factor of $1/4\pi$ in
\eref{e:muu_def}.  Later, the Hellings and Downs curve was
reinterpreted as a {\it correlation coefficient} for the redshift
response of two pulsars, responding to the same GW source. 
For this case, one wants the correlation coefficient to equal unity
if the two pulsars are identical (i.e., $a=b$).
Since the expected correlation for two identical pulsars is twice 
that for two {\it distinct} pulsars separated by $0^\circ$
[due to the $\delta_{ab}$ term in \eref{e:source_avg}], one needs 
$\muu(0)=1/2$ for the correlation-coefficient interpretation.
This means multiplying the right-hand sides of \eref{e:muu_def}
and \eref{e:HD} by 3/2.

\vspace{0.1in}\hrule
\bi
\item[Q2:] 
Why does the Hellings and Downs curve have different values at
$0^\circ$ and $180^\circ$ if the quadrupolar deformation of space
produced by a passing GW affects two test masses $180^\circ$
apart in exactly the same way?
\ei

\begin{figure}[htbp]
\begin{center}
\includegraphics[width=\textwidth]{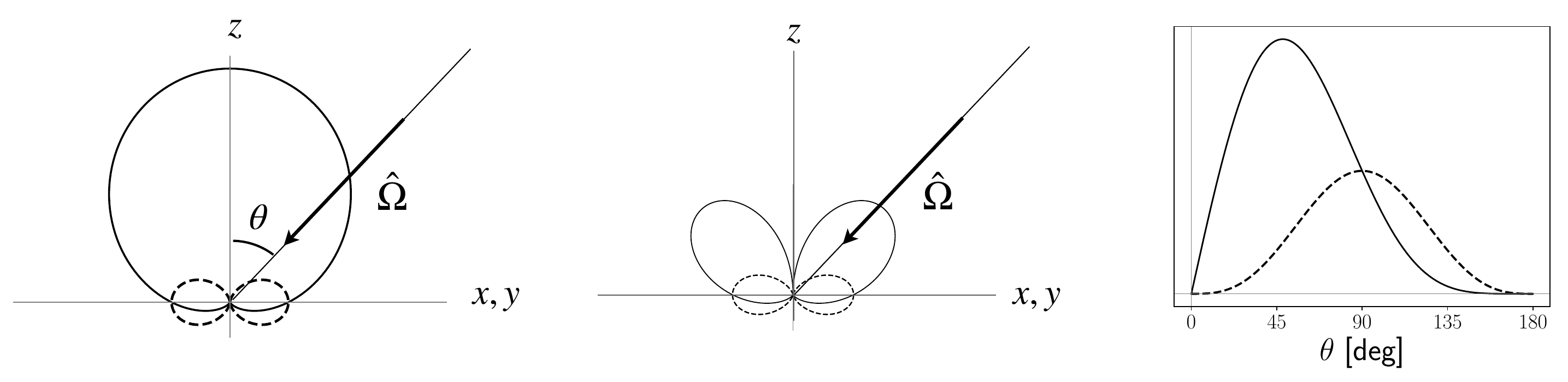}
\caption{Integrands of the Hellings and Downs correlation curve
for two pulsars separated by $0^\circ$ ($\hat p_1=\hat p_2=\hat z$, 
solid curve) and by $180^\circ$ ($\hat p_1=-\hat p_2=\hat z$, 
dashed curve).  The first two panels are polar
plots, which are axially symmetric about the $z$-axis.  The left
plot shows
$F_1^+(\Omg)F_2^+(\Omg) + F_1^\times(\Omg)F_2^\times(\Omg)$
for the two different choices of $\hat p_1$ and $\hat p_2$.
The polar plots in the middle panel include an additional
factor of $\sin\theta$, which arises when integrating over
$\theta$. The right panel is the same as the middle panel, 
but plotted as a function of $\theta$.
The area under the 
solid curve is twice the area under the dashed curve.}
\label{f:overlapComparison_0_180}
\end{center}
\end{figure}

\noindent
Answer: 
The Hellings and Downs curve has different values
at $0^\circ$ and $180^\circ$ due to the different 
denominators $1+\Omg\cdot\hat p_1$ and 
$1+\Omg\cdot\hat p_2$,
which appear in the antenna pattern functions 
$|F_1(\Omg)|$ and $|F_2(\Omg)|$ for the
two pulsars.
The two cases to consider for pairs of pulsars 
separated by $0^\circ$ and $180^\circ$ are 
$\hat p_1=\hat p_2=\hat z$ and 
$\hat p_1=-\hat p_2=\hat z$, respectively.
This difference is illustrated graphically in the 
left panel of \Fref{f:overlapComparison_0_180}, where
we plot the integrand
$F^+_1(\Omg) F^+_2(\Omg) + F^\times_1(\Omg) F^\times_2(\Omg)$ 
of the expected Hellings
and Downs correlation for these two different cases.
It is apparent from this plot that if the GWB consisted
of waves that only came from directions 
{\it perpendicular} to the Earth-pulsar baselines (so, in
the $xy$-plane of this plot), then the values of the 
Hellings and Downs correlation at $0^\circ$ and $180^\circ$
would be exactly the same.
But since an isotropic GWB has equal contributions 
from GWs coming from {\it all} different directions on the sky, 
the Hellings and Downs correlation for the case 
$\hat p_1=\hat p_2$ will be larger than that for 
$\hat p_1=-\hat p_2$.
Why these values differ by exactly a factor of two is 
the topic of the next question.

\vspace{0.1in}\hrule
\bi
\item[Q3:] 
Why is the value of the Hellings and
Downs curve for two pulsars separated by $180^\circ$ 
exactly half that for $0^\circ$ angular separation?
\ei

\noindent
Answer: To prove this, 
start with~\eref{e:muu_def} for the expected Hellings
and Downs correlation and~\eref{e:F^A} for the 
Earth-term-only response functions $F^A(\Omg)$.
Taking the two pulsars to point in the same direction
($\hat p_1=\hat p_2 = \hat z$), we have
\be
\sum_A F_1^A(\Omg) F_2^A(\Omg) = \frac{1}{4}(1+\cos\theta)^2 = \frac{1}{4}(1+u)^2\,,
\label{e:F1F1}
\ee
where $u = \cos \theta$. 
Having them point in opposite directions
($\hat p_1 = -\hat p_2=\hat z$) leads to
\be
\sum_A F_1^A(\Omg) F_2^A(\Omg) = \frac{1}{4}(1-\cos^2 \theta)
= \frac{1}{4}(1-u^2)\,.
\label{e:F1F2}
\ee
These functions are plotted in \Fref{f:overlapComparison_0_180}, where
$\theta$ is the usual polar angle measured with respect to the
$z$-axis.  [Note: the comment following \eref{e:n} explains why
$u=\cos \theta$ has the opposite sign in some of the literature, for
example in Eq.~D10 of~\cite{AllenHDVariance}.]

The overlap function is the average of this quantity over the sphere.
From the three
panel plots in \Fref{f:overlapComparison_0_180}, we see that when the
two pulsars both point in the $\hat z$ direction, the majority of
support for the overlap function comes from sky directions $\hat
n=-\Omg$ having $z>0$.  When the two pulsars point in opposite
directions, $\hat p_1=-\hat p_2=\hat z$, the overlap function has
equal contributions from $z>0$ and $z<0$.  These contributions are
only slightly larger than those for $\hat p_1=\hat p_2$ for $z<0$, but
much smaller for $z>0$.

Visual inspection of the areas under the two curves shown in the
rightmost panel of \Fref{f:overlapComparison_0_180} shows why
the value of the HD curve for two pulsars separated by $0^\circ$ is
twice as large as that for two pulsars separated by $180^\circ$.
Indeed, 
averaging these quantities
over the sphere gives
\be
\muu(0^\circ) = \frac{1}{3}\quad(\hat p_1=\hat p_2=\hat z)\,,
\quad
\muu(180^\circ) = \frac{1}{6}\quad(\hat p_1=-\hat p_2=\hat z)\,.
\ee
These values are easy to check, since the average of a
function $Q(u)$ over the sphere is $(1/4 \pi) \int d\Omg\> Q(\cos
\theta) = (1/4 \pi) \int_0^{2 \pi} \d\phi \int_0^\pi \d \theta\>
\sin \theta\, Q(\cos \theta) = (1/2) \int_{-1}^1 \d u\>Q(u)$. The
integrals of \eref{e:F1F1} and \eref{e:F1F2} with respect to $u$ are
trivial.

\vspace{0.1in}\hrule
\bi
\item[Q4:] 
Why is the most negative value of the Hellings and Downs 
curve not at $90^\circ$?
\ei

\noindent
Answer: 
Again, this is because the Earth-term-only
response functions $F^A(\Omg)$ given in \eref{e:F^A}
are not symmetric under $\hat p\rightarrow - \hat
p$.  If the Hellings and Downs correlation were purely quadrupolar,
then its most negative value would be exactly at $90^\circ$, as it
is for the case of an array of short-arm LIGO-like detectors
(see the right panel of \Fref{f:HD_P2}).  If it
had only even multipoles (no odd-$l$ terms), then it would be
reflection symmetric about $90^\circ$, but it might not have its minimum
there.
  
Although the dominant contribution to the Hellings and Downs
curve is the quadrupole pattern $P_2(\cos\gamma)$, there are
nonzero contributions from all higher order $l$-modes as well.  We
leave it as an exercise for the reader to show that
\begin{eqnarray}
  \label{e:HD_legendre2}
\muu(\g) 
= \sum_{l=2}^{\infty}
(2l+1) c_l P_l(\cos\gamma)\,,
\quad{\rm where}
\\c_l = [(l+2)(l+1)l(l-1)]^{-1}\,. 
\label{e:HD_legendre}
\end{eqnarray}
To obtain this result, one needs to use (a) the orthogonality property
\be
\int_{-1}^1 \d x\> P_l(x) P_{l'}(x) = \frac{2}{2l+1}\delta_{ll'}
\ee
of the Legendre polynomials, where $x\equiv \cos\gamma$, (b) Rodrigues' Formula
for Legendre polynomials, and (c) integration by parts.  See equations
(97)--(105) in~\cite{PhysRevD.90.082001} for a step-by-step
derivation of~\eref{e:HD_legendre2} and~\eref{e:HD_legendre}, 
or see Secs.~III and IV of \cite{allen2024angular}.

\begin{figure}[htbp]
\begin{center}
\includegraphics[width=.6\textwidth]{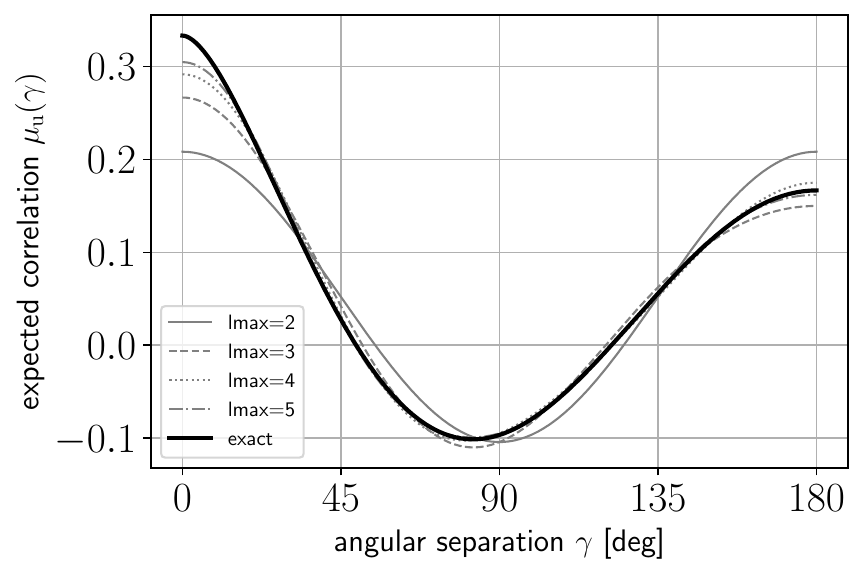}
\caption{Approximations to the Hellings
and Downs curve (solid black)
obtained by terminating the sum in 
\eref{e:HD_legendre} with two to five Legendre polynomials 
(grey curves).}
\label{f:HD_approx}
\end{center}
\end{figure}

\Fref{f:HD_approx} compares approximate versions of
the Hellings and Downs curve, terminating the sum at
different maximum values 
$l_{\rm max}=2,3,4,5$.
Since the $c_l$'s in~\eref{e:HD_legendre} fall off like
$1/l^4$, only a few values of $l$ are needed to get a very good 
approximation.

\vspace{0.1in}\hrule
\bi
\item[Q5:] Why is the Hellings and Downs curve/correlation frequency
  independent, whereas overlap functions for Earth-based
  interferometers are frequency dependent, for example, as shown in
  Figure~2 of Ref.~\cite{AllenRomano}?
\ei

\noindent
Answer: The expected Hellings and Downs correlation for a 
pair of pulsars $a$ and $b$ does not depend on frequency
since the frequency dependence of the redshift response
functions $R^A_a(f,\Omg)$ and $R^A_b(f,\Omg)$ 
given in \eref{e:R^A} enters only via the pulsar terms 
$e^{-i2\pi fL_a(1+\Omg\cdot\hat p_a)}$ and
$e^{-i2\pi fL_b(1+\Omg\cdot\hat p_b)}$.
These terms drop out when calculating the expected correlation 
of $Z_a(t)$ and $Z_b(t)$, as described in \Sref{s:source_averaging}.

However, for a pair of Earth-based interferometers, each of which has
two perpendicular equal-length arms, the corresponding response
functions have the form~\cite{Flanagan:1993}:
\bnp
\bea
R_a^A(f,\Omg)=F_a^A(\Omg)e^{-i2\pi f\Omg\cdot\vec r_a}\,,
\label{e:R^A_IFO}
\\
F_a^A(\Omg)\equiv \frac{1}{2}(\hat X_a^i\hat X_a^j-\hat Y_a^i\hat Y_a^j)e^A_{ij}(\Omg)\,,
\quad A=+,\times\,,
\label{e:F^A_IFO}
\eea
\enp
where $\hat X_a$, $\hat Y_a$ are unit vectors pointing along the 
two arms of the interferometer, and 
$\vec r_a$ is the position vector of the interferometer's central station.
(Similar equations hold for the second interferometer with $a$ replaced
by $b$.)
When forming the expected correlation, we need to evaluate the integral
\be
\frac{1}{4\pi}\int {\rm d}\Omg\>
\sum_A F^A_a(\Omg) F^A_b(\Omg)\, e^{-i2\pi f\Omg\cdot(\vec r_a-\vec r_b)}\,.
\ee
Now, for two interferometers on Earth, the distance $D_{ab}\equiv |\vec r_a-\vec
r_b|$ between the central stations is comparable to the wavelength
$\lambda\equiv 1/f$ (hundreds to thousands of km) of the GWs that
these instruments detect.  Hence, this exponential factor 
cannot be approximated by unity.
(As discussed in Sec.~\ref{s:intro},
for PTAs, the timing residuals are determined at
the same point, the SSB, for all pulsars and all telescopes.
So for PTAs,
$\vec r_a-\vec r_b=\vec 0$, implying that $e^{-i 2\pi f\Omg\cdot(\vec r_a-\vec r_b)}=1$.)
Thus, while individual Earth-based
interferometers operate in the short-arm limit $L\ll \lambda$, 
it is {\it not} also the case that $D_{ab}\ll\lambda$, but rather that
$\lambda\lesssim D_{ab}$. Since the frequency dependence for
Earth-based interferometers is tied to the detector locations, it
cannot be factored out of the correlation as a detector-independent
function of frequency.

\vspace{0.1in}\hrule
\bi
\item[Q6:] 
Does recovery of the Hellings and Downs curve from observed pulsar-pair correlations imply that
the GWB is isotropic?
\ei

\noindent
Answer: 
No.  The Hellings and Downs curve is a robust
prediction for the correlation produced by any (small or large) set
of noninterfering GW sources, provided pulsar averaging is 
carried out using many pairs of pulsars uniformly distributed over the sky.  
It is {\it independent} of the angular distribution of the GW sources 
on the sky.
This result was first demonstrated for the special case 
of GWs produced by a single black-hole binary in~\cite{Cornish_2013}.  
It was then extended to an arbitrary set of noninterfering 
sources in \cite{AllenHDVariance} (see also FAQ Q7 below), 
which further exploited pulsar averaging.  
For interfering sources, we do {\it not} expect to recover 
exactly the Hellings and Downs curve (see \cite{AllenHDVariance} and FAQ Q7 
below).

As discussed in \Sref{s:source_averaging},
the original/traditional approach to compute the expected
Hellings and Downs correlation is to take a single fixed 
pair of pulsars, and then to average their correlation
over GW propagation directions $\Omg$, assuming that the sources are 
unpolarized and isotropic on the sky~\cite{HD},
and that they do not interfere with each other.
In references~\cite{Cornish_2013} and \cite{AllenHDVariance}, 
somewhat involved calculations were done to show the 
equivalence of pulsar averaging and source averaging 
for the case of a single GW source, see \eref{e:G=muu}.

Happily, one can show the equivalence of these two
approaches without any calculation. Here, we show that pulsar
averaging is equivalent to GW-source averaging with a purely
geometric argument, using a series of three Euler-angle
rotations. At first, this appears to require adding something to the
traditional source-averaging calculation: in addition to averaging
over the GW propagation direction $\Omg=\Omg(\theta,\phi)$, one must
\emph{also} average over the polarization angle $\psi$.  In fact,
while often overlooked, this is \emph{also} required by the
traditional calculation: it gives rise to the quadrature sum.  It is
overlooked because the original~\cite{HD} calculation obscures this
point by not giving enough detail.  To obtain that result, one must
assign a polarization direction or angle to each source, and also
average over that, for example, as in the calculation starting from
\eref{e:redshiftUnpolSource2}.  Hence, even in the traditional
GW-source averaging approach, the averages must be calculated with
respect to three angles, not just two.

\begin{figure}[htbp]
\begin{center}
\includegraphics[width=.8\textwidth]{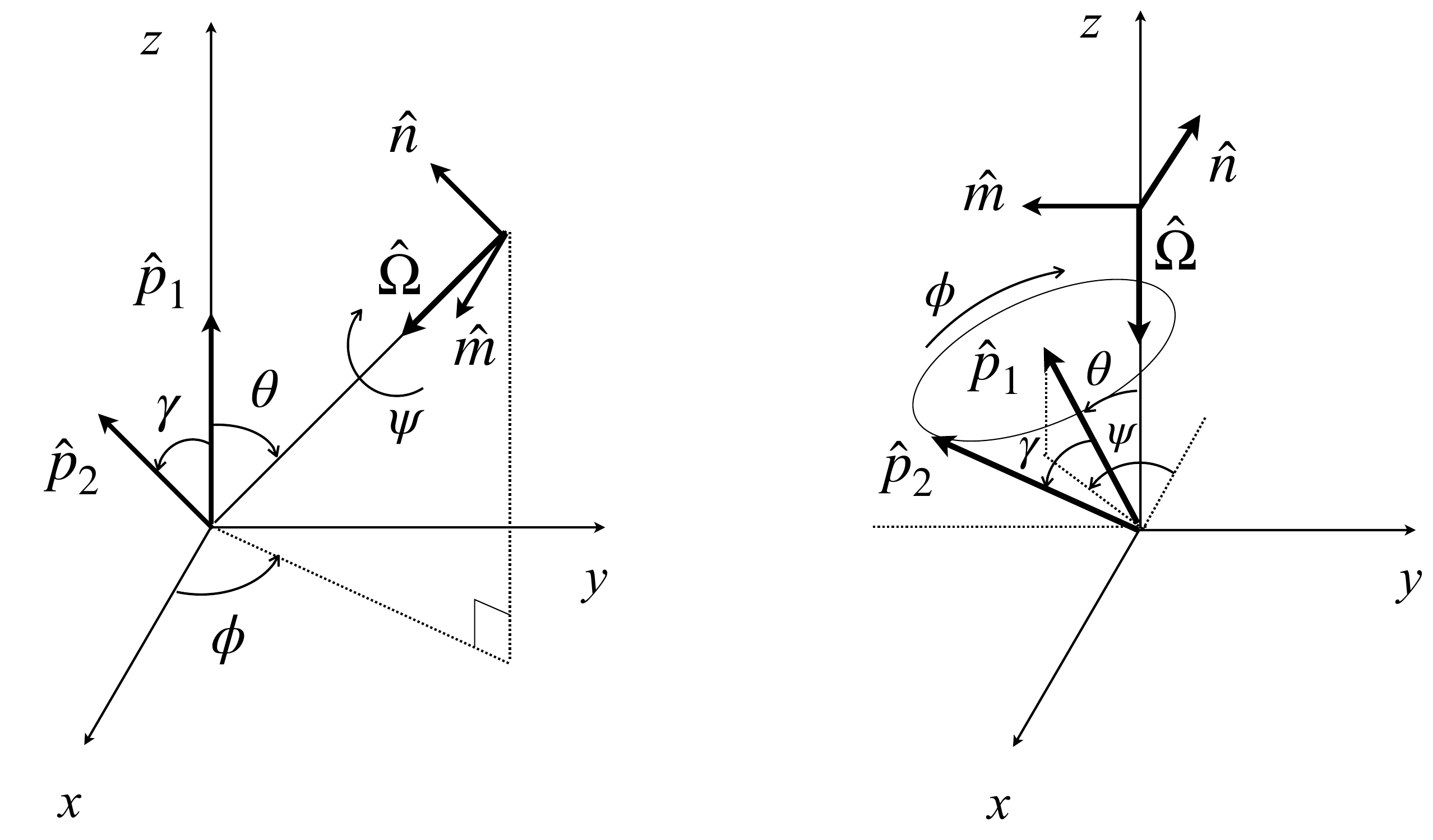}
\caption{
GW source averaging (left panel) and pulsar averaging 
(right panel) are equivalent, as can be seen by carrying
out a series of rotations.}
\label{f:pulsar_GW_averaging}
\end{center}
\end{figure}

To see the equivalence purely from geometry, start with the averaging over 
GW source directions. As previously done, without loss of generality, put
$\hat p_1$ along the $z$-axis and $\hat p_2$ in the $xz$-plane, so
that $\hat p_2= \sin\gamma\,\hat x +\cos\gamma\,\hat z$.  In contrast, pulsar
averaging is carried out with respect to the angular coordinates of
$\hat p_1$ (polar coordinate $\theta$, azimuthal coordinate $\psi$)
and those of $\hat p_2$ (which lies on a cone about $\hat p_1$ with
fixed half-opening angle $\gamma$, with its location on the cone parameterized
by $\phi\in[0,2\pi)$).  See \Fref{f:pulsar_GW_averaging} and
compare with Figure~5 of~\cite{AllenHDVariance}.

The sequence of rotations needed to go from source 
averaging to pulsar averaging is as follows: 
first rotate by $-\phi$ about the $z$-axis, which
moves the GW propagation direction $\Omg$ into the 
$xz$-plane and $\hat p_2$ around a cone with 
central axis $\hat p_1$ through angle $\phi$.
Then rotate by $-\theta$ around the $y$ axis, which
puts $\Omg$ along $-\hat z$, and makes $\hat p_1$ have 
polar angle $\theta$.
Finally, rotate around $\Omg$ (which now points
in the $-\hat z$ direction) by $-\psi$, which 
becomes the azimuthal angle for $\hat p_1$ (with respect
to the $-x$-axis).

In practice, we do not have an infinite number of pulsar pairs
separated by the angle $\gamma$. So, to do pulsar averaging, one
divides the $N_{\rm pul}(N_{\rm pul}-1)/2$ interpulsar correlations
into a smaller number $N_{\rm bins}$ of angular separation bins.  This
form of pulsar averaging is always used when attempting to recover or
reconstruct the Hellings and Downs curve based on observed interpulsar
correlations.  For example, for the NANOGrav 15-year analysis shown in
\Fref{f:HD-NG}, $N_{\rm pul}=67$ and $N_{\rm bins}=15$, which implies
that the estimated correlation value in each angular separation bin
involved averaging the correlations of approximately 150 pulsar pairs.

\vspace{0.1in}\hrule
\bi
\item[Q7:]
 In the distant future, when PTA observations are carried
  out with larger and more sensitive telescopes, more pulsars, and
  much longer observation periods, will we {\it exactly} recover the
  Hellings and Downs curve?
\ei

\noindent
Answer:
As we shall show below, this would be the case if the GWB were produced
by a set of discrete noninterfering 
sources~\cite{AllenHDVariance, AllenFrascati:2022}.
But it is much more likely that our Universe contains many GW sources
radiating at frequencies that are close enough together that they 
cannot be resolved individually over the time span of the PTA
experiments (say, tens to hundreds of years).  This means that the
GWB is produced by a sum of {\it interfering sources}.  An example
of such a GWB is the ``confusion noise" created by a set of
sources randomly distributed on the sky,
emitting monochromatic GWs with random
phases,
but occupying {\it the same frequency bin} \cite{allen2024pulsar}.  In this case,
pulsar-averaged noise-free measurements over an infinite number of
pulsar pairs with the same angular separation do not give exactly the
Hellings and Downs curve.

If we have an infinite collection of universes, each of which has GW
sources having independent random phases, and we were also able to
average the correlation over all of those universes (called an
``ensemble average"), then we would get exactly the Hellings and Downs
curve.  But in any given universe, we would find something different
from Hellings and Downs.  The (mean squared) difference between what
would be observed in any given universe, and the (unobservable)
average over many universes, is called the {\it cosmic variance}.
In what follows, we show how the cosmic variance arises from interference
between different GW sources.

Consider a universe which contains $N$
point sources of GWs, which we label $n=1, \dots, N$.  These GW
sources are located far from the neighborhood containing the Earth and
the PTA pulsars.  The GW sources have sky locations $-\Omg_n$, and the
$n$'th source produces a complex strain at Earth $h_{ij}(t) = \Re
[h_n(t) e^*_{ij}(\Omg_n)]$, where the complex waveform is $h_n(t) =
h_n^+(t) + i h_n^+(t)$.  Since the PTA is far from the sources, these
GW strains are small, so the total strain at Earth can be obtained by
summing them.  The redshift of pulsar $a$ is then obtained by summing
\eref{e:Z(t)} to obtain
\bnp 
\bea
\label{e:reda}
Z_a(t) &= \Re\left[\sum_{n=1}^N h_n(t) F^*_a(\Omg_n)\right]
\\
&= \frac{1}{2} \sum_n \Bigl[ h_n(t) F^*_a(\Omg_n) +  h^*_n(t) F_a(\Omg_n) \Bigr]\,,
\label{e:redb}
\eea
\enp
where $F_a(\Omg) \equiv F_a^+(\Omg) + i F_a^\times(\Omg)$ are the
complex antenna patterns, and we adopt the convention that sums over
$n$ and $m$ are from $1,\cdots, N$ unless otherwise indicated.  A
similar expression holds for pulsar $b$.  We assume that the pulsars
$a$ and $b$ are distinct and are separated from one another by many GW
wavelengths, and therefore have neglected the pulsar terms for the
reasons discussed earlier.  (We work with complex quantities because
it simplifies some of the following expressions.)

The correlation between the redshifts of the 
two pulsars is the time-averaged product
$\rho_{ab} \equiv \overline{Z_a(t) Z_b(t)}$, which evaluates to
\bea
\fl
\hspace{0.35in}
\rho_{ab} 
= \frac{1}{4} \sum_n \sum_m 
\Bigl[\,\overline{h_n h_m^*} F^*_a(\Omg_n)F_b(\Omg_m) 
+  \overline{h^*_n h_m} F_a(\Omg_n)F^*_b(\Omg_m)  
\nonumber\\
\hspace{0.65in}
+ \overline{h_n h_m} F^*_a(\Omg_n)F^*_b(\Omg_m) 
+ \overline{h^*_n h^*_m} F_a(\Omg_n)F_b(\Omg_m) \Bigr]\,.
\label{e:rhoab}
\eea
To make the expression more compact, we do not explicitly show the 
time dependence of the quantities being time-averaged.

Consider the pulsar average of~\eref{e:rhoab}, which is
obtained by averaging the correlation over all pulsar pairs $ab$
separated by angle $\g$.  As discussed briefly in
Sec.~\ref{s:pulsar_averaging},
and in more detail in \cite{AllenRomanoOptHD},
this corresponds closely to
experimental practice: as illustrated in figure~\ref{f:HD-NG},
PTA experiments ``reconstruct'' the
angular dependence of the Hellings and Downs correlation via a
pulsar averaging procedure. Also, note that we dropped the pulsar
terms from \eref{e:reda}.  Had we included them, they would have
been eliminated by this pulsar averaging step, as shown in Sec.~IV
of \cite{AllenHDVariance}.

Consider first the case where there is no interference between GW
sources, for example, because the sources are radiating GWs at
different frequencies.  This means that the time averages 
$\overline{h_n h_m}$ and $\overline{h_n h^*_m}$ vanish if $n$
is different than $m$. Thus, in this no-interference case, the double
sum reduces to a single sum:
\bea
\fl
\hspace{0.35in}
\rho_{ab} =  \frac{1}{4} \sum_n \Big[\,
  \left(F^*_a(\Omg_n)F_b(\Omg_n) +  F_a(\Omg_n)F^*_b(\Omg_n) \right)  \overline{|h_n|^2}
\nonumber\\
\hspace{1.65in}
+ 2 \Re \left( F^*_a(\Omg_n)F^*_b(\Omg_n) \, \overline{h^2_n} \right)  \Big].
\label{e:rhoabNoI}
\eea
We now pulsar-average the correlation~\eref{e:rhoabNoI}.  The only
dependence on $a$ and $b$ in~\eref{e:rhoabNoI} is via the antenna
pattern functions.  As explained in Q6 and
\Fref{f:pulsar_GW_averaging}, the pulsar average of $F_a(\Omg_n)
F^*_b(\Omg_n)$ is \emph{independent of the source direction} $\Omg_n$
and gives exactly the Hellings and Downs curve $\muu(\g)$.  The pulsar
average of $F_a(\Omg_n)F_b(\Omg_n)$ vanishes.  Hence, the pulsar
average of \eref{e:rhoabNoI} gives
\be
\label{e:rhoabNoI2}
\langle\rho_{ab}\rangle_{ab\in\g} =  
\frac{1}{2} \Bigl( \sum_n \overline{|h_n|^2} \Bigr) \muu(\g)\,,
\ee
where the subscript ``$ab\in\gamma$'' indicates that we have
averaged over pulsar pairs separated by angle $\gamma$.  Thus, we have
proved that if there is no interference between the GW sources, then
the pulsar average of the correlation exactly matches the Hellings and
Downs curve {\it in any instance of the universe}.

However, the situation changes in an important way if the GW sources
interfere.  This seems to be a possibility that Ron Hellings and
George Downs had not considered, perhaps because they approached the
topic from the perspective of radio astronomy.  In radio astronomy,
distinct radio sources, observed over a long period of time at
different sky locations, undergo destructive interference; in our
language they do not interfere.  (At GHz frequencies, these sources go
through $10^9$ oscillations per second, and over intervals longer than
a small fraction of a second, the time-averaged product of their
electric field components is zero.)  However, for PTAs, the relevant
GW sources undergo at most a few oscillations over an observation
interval of years or decades.  Although the sources have no causal
connection, and are ``independent'' in the normal sense, the
time-averaged product of their strain waveforms can nevertheless have
a magnitude which is comparable to the time-averaged square of either
one. The sign is positive if the two sources happen to be in phase,
negative if they happen to be out of phase, and zero if two sources
are exactly in quadrature.  Thus, the case of interfering GW sources
is not just realistic, it is inevitable for the sources relevant to
PTAs for the typical PTA observation times.

If the GW sources interfere, then the pulsar average of~\eref{e:rhoab}
contains the ``diagonal'' $n=m$ terms whose pulsar average we found
in~\eref{e:rhoabNoI2}, and also has additional contributions arising
from the off-diagonal $n\ne m$ interference terms.  To
  understand the effects of these interference terms, it is helpful to
  introduce the ``Hellings and Downs two-point function''.  This was
  first defined and computed for a particular pair of GW source
  directions in Appendix~G of~\cite{AllenHDVariance}.  It was then
  generalized to arbitrary source directions
  in~\cite{allen2024angular}, giving rise to an additional
  unit-modulus complex phase, whose argument $2 \chi$ is defined
  by~\cite[Eq.~(6.5)]{allen2024angular}.

The Hellings and Downs two-point function is the pulsar average of
$F_a(\Omg_n)F^*_b(\Omg_m)$ over all pulsar pairs $a$ and $b$,
uniformly distributed over the sphere, and separated by angle
$\gamma$, as discussed in Sec.~\ref{s:pulsar_averaging}.  It may be
written in the form 
\be
\label{e:twopointfn}
\langle F_a(\Omg_n)F^*_b(\Omg_m) \rangle_{ab \in \gamma} =
\mu(\gamma, \beta_{nm}) {\rm e}^{2 i \chi^{\phantom{'}}_{nm}}
\, ,
\ee
where $\mu(\gamma, \beta)$ is a real function of two variables given
explicitly in Eq.~(G5) of~\cite{AllenHDVariance} and plotted in
Figure~12 of that reference.  We will assume that the reader is
familiar with this plot; $\gamma$ is the angle between
the directions to the two pulsars in the averaging ($\cos \gamma = \hat p_a \cdot \hat
p_b$), and $\beta_{nm} = \beta_{mn}$ is the angle between the
directions to the two GW sources ($\cos\beta_{nm} = \Omg_n \cdot
\Omg_m$).  The real angle $\chi^{\phantom{'}}_{nm} \equiv \chi(\Omg_n,\Omg_m)$,
where $\chi(\Omg_n,\Omg_m)$ is defined by~\cite[Eq.~(6.5)]{allen2024angular}.
Note that $\chi$ is an antisymmetric
function of the positions $\Omg_n$ and $\Omg_m$ of the 
sources\footnotemark{} $\chi^{\phantom{'}}_{nm} = -\chi^{\phantom{'}}_{mn}$;
it vanishes for the particular GW source locations $\Omg$ and $\Omg'$ considered
in~\cite{AllenHDVariance}.  Note that if the two GW sources are
coincident, then $\beta_{nm} = \chi^{\phantom{'}}_{nm}=0$, and the two-point
function reduces to $\mu(\gamma,0) = \muu(\g)$, which is the Hellings
and Downs curve.

The pulsar average of $F_a(\Omg_n)F_b(\Omg_m)$ can be obtained from
\eref{e:twopointfn} by sending both $b$ and $\Omg_m$ to their
antipodal points on the sphere, as detailed in~\cite{AllenHDVariance,
  allen2024angular}.  This sends $\gamma \to \pi - \gamma$ and $\beta
\to \pi - \beta$ and also modifies the phase; one obtains\footnotemark[\value{footnote}] 
\be
\langle F_a(\Omg_n)F_b(\Omg_m) \rangle_{ab \in \gamma} =
\mu(\pi-\gamma, \pi-\beta_{nm}) {\rm e}^{2 i \chi'_{nm}} \,,
\ee
where $\chi'_{nm} \equiv \chi(\Omg_n,-\Omg_m)$ is real and symmetric, so $\chi'_{nm} = \chi'_{mn}$; 
see~\cite[Eq.~(C4)]{allen2024angular}.
If the two GW
sources are antipodal, then the two-point function vanishes:
$\mu(\gamma, \pi) = 0$, as can be seen in Figure~12
of~\cite{AllenHDVariance}.
\footnotetext{While this analysis only needs the
    symmetry properties of $\chi_{nm}$ and $\chi'_{nm}$, for
    completeness we note that $\chi^{\phantom{'}}_{nm} \equiv
    \chi(\Omg_n,\Omg_m)$ and $\chi'_{nm} \equiv \chi(\Omg_n,-\Omg_m)$,
    where $\chi(\Omg,\Omg')$ is defined
    by~\cite[Eq.~(6.5)]{allen2024angular}.  In this definition of
    $\chi'$, the second argument has been shifted to the antipodal
    point.  Thus, if $\Omg_m$ has spherical polar coordinates
    $(\theta_m,\phi_m)$ then its antipodal point  $-\Omg_m$ 
    has coordinates $(\pi-\theta_m,\phi_m + \pi)$.}

Thus, in the case where the GW sources interfere, the pulsar average
of~\eref{e:rhoab} gives
\bea
\label{e:rhoabNoI3}
\fl \langle\rho_{ab}\rangle_{ab\in\g} & =  & \frac{1}{2} \Bigr( \sum_n \overline{|h_n|^2} \Bigr) \muu(\g)
+ \\   \nonumber \fl
&  & \frac{1}{2} \Re \sum_{n \ne m}  \Bigl[ 
   \bigl(\overline{h^*_n h_m}  {\rm e}^{2i\chi^{\phantom{'}}_{nm}} \! \bigr) \mu(\gamma, \beta_{nm}) +
  \bigl( \overline{h^*_n h^*_m}    {\rm e}^{2i\chi'_{nm}} \! \bigr) \mu(\pi-\g, \pi-\beta_{nm}) \Bigr]\,.
\eea
The key point is this: the additional $N(N-1)$ terms arising from
the two-point function, considered as functions of the angle $\gamma$,
are \emph{not proportional to the Hellings and Downs curve}
$\muu(\g)$.  Moreover, if the number of sources is large, then these
terms make contributions to the pulsar average which are comparable to
the first (diagonal) term.

We demonstrate this for the simple ``confusion-noise model''
constructed in~\cite{AllenHDVariance}. In this model, each of the $N$
sources of GWs is circularly polarized, and all of them radiate at the
same GW frequency $f$, which is picked to be commensurate with the PTA
observation interval.  From \eref{e:h_circ}, the waveforms are
\be
h_n(t) = A_n \exp[i (2\pi f t + \phi_n)]\,,
\label{e:simplemodel}
\ee
where the amplitudes $A_n > 0$ are real, the GW frequency $f$ is the
same fixed multiple of $1/T$ for all $N$ sources, and the phase
$\phi_n$ of the $n$'th source is a real number in the interval
$[0,2\pi)$.  For this model, the time averages which appear in
  \eref{e:rhoabNoI3} are $\overline{h^*_n h_m} = A_n A_m \exp[i(\phi_m
    - \phi_n)]$ and $\overline{h^*_n h^*_m} = 0$.  Thus, the
  pulsar-averaged correlation~\eref{e:rhoabNoI3} simplifies to
\be
\fl
\hspace{0.5in}
\langle\rho_{ab}\rangle_{ab\in\g} =  \frac{1}{2} \Bigl( \sum_n  A^2_n \Bigr) \muu(\gamma) 
+ \frac{1}{2} \sum_{n\ne m} \! A_n A_m \cos \Delta_{nm} \, \mu(\gamma, \beta_{nm})\,,
\nonumber
\label{e:rhoabNoI4}
\ee
where $\Delta_{nm} \equiv \phi_n - \phi_m - 2 \chi^{\phantom{'}}_{nm}$ is a real
phase.  Because the ``cross-sections'' of $\mu(\gamma, \beta)$ for
$\beta \ne 0$ do not have exactly the same shape as the HD curve, this
means that in any given realization of the universe, the
pulsar-averaged correlation will not match the HD curve exactly.  (If
we could take the correlation curves from many different random
realizations of the universe and average them together, then the terms
from the second sum would average to zero, and we would obtain exactly
the HD curve.)

It is easy to quantify the amount by which a typical universe will vary away from
the average of many universes, which is the HD curve.
Consider how
the $N(N-1)$ terms of the second sum add together.  If the phases
$\phi_n$ of the $N$ sources are random and independent, then the
cosine has values that lie in the interval $[-1,1]$, so the sum
includes both positive and negative terms.  If the number of GW
sources $N$ is large, then this second sum can be thought of as doing
a random walk with $N(N-1)$ steps to the left and right along the real
number line.  It thus contributes a value whose typical magnitude is
of order $\sqrt{N(N-1))} \approx N$, which is comparable to the
magnitude of the first term.  So, even if the number of sources $N \to
\infty$, the second term gives rise to a nonzero deviation away from
the Hellings and Downs curve.  The mean-squared value of this
deviation is exactly the cosmic variance.

We can easily estimate the size of the cosmic variance, starting from
\eref{e:rhoabNoI4}.  The mean value, assuming independent random
phases, and averaging over many different universes, is the first term
in \eref{e:rhoabNoI4}.  The deviation away from that (unobservable)
mean, in any given realization of the universe, is given by the second
term in \eref{e:rhoabNoI4}.  If the GW sources are uniformly
distributed over the two-sphere, then the distribution of the angles
$\beta_{nm}$ is proportional to $\sin \beta_{nm}$.  Thus, the expected
squared deviation of the pulsar-averaged correlation away from the
Hellings and Downs curve, as a function of $\gamma$, has the
form%
\footnote{The $\chi^{\phantom{'}}_{nm}$ do not affect the
  distribution over $\beta$ in \eref{e:rhoabNoI5}: they are eliminated
  from the second moment when $\cos \Delta_{nm} \cos \Delta_{kl}$ for
  $n \ne m$ and $k \ne l$ is averaged over the random phases
  $\phi_j$. To see this, write the cosines as exponentials. The
  average has four terms.  A typical term is ${\rm e}^{i(\Delta_{nm} +
    \Delta_{kl})}$, whose random phase average is $\delta_{nl}
  \delta_{mk} {\rm e}^{-2i (\chi^{\phantom{'}}_{nm} +
    \chi^{\phantom{'}}_{kl})}$.  The $\chi$'s cancel and drop out because
  $\chi^{\phantom{'}}_{nm} + \chi^{\phantom{'}}_{mn}$ vanishes.}
\be
\label{e:rhoabNoI5}
\sigma^2_{\rm cosmic} (\gamma) \propto \int_0^\pi \d\beta\>
\sin \beta  \, \mu^2(\gamma, \beta) \,\,
\, = 2 \sum_{l=2}^{\infty}
(2l+1) c^2_l P^2_l(\cos\gamma)
\,.
\ee
In the final equality, $c_l$ are defined by \eref{e:HD_legendre};
see~\cite[Eq.~(8.3)]{allen2024angular}.
For a more realistic set of GW sources, the full formula is derived
in~\cite{AllenHDVariance, AllenRomanoOptHD, AllenFrascati:2022}.  The basic
ideas are the same, so it is not surprising that the equation is identical to
\eref{e:rhoabNoI5}, except that it also includes the overall coefficient, which
we have omitted.  In particular, the angular dependence on $\gamma$ is exactly
as given in \eref{e:rhoabNoI5}, with an overall coefficient that correctly
weights the contributions of GW sources radiating at different frequencies.

\vspace{0.1in}\hrule
\bi
\item[Q8:] 
Assuming noise-free observations, are the fluctuations
away from the expected Hellings and Downs curve only due to 
the pulsar-term contributions to the timing residual response?
\ei

\noindent
Answer:
No. While the pulsar terms do contribute to
deviations away from the expected Hellings and Downs correlation,
there are other more fundamental sources of such deviations.  (i)
Timing residual or redshift measurements from two different pulsar pairs
separated by the same angle $\gamma$ but pointing in different
directions on the sky will have different correlations due to the different
parts of the GWB that they are sensitive to.
So, if the power in the GW signal is not uniformly distributed, 
this gives deviations.
(ii) For interfering sources, even with infinite pulsar averaging 
as described in FAQ Q7, there will be fluctuations away from the
expected Hellings and Downs correlation that are associated with the
Earth-term-only response functions.

For a given pulsar pair, the pulsar-term contributions to the timing
residual response {\it add} to the fluctuations, with the same
root-mean-square amplitude as the Earth terms.  But, as mentioned in 
the previous paragraph, they are not the
sole source of the fluctuations away from the Hellings and Downs
curve, and as more and more pairs of pulsars are averaged
together at a particular angular separation, this pulsar-term
contribution becomes insignificant.

\vspace{0.1in}\hrule
\bi
\item[Q9:] Why does the Hellings and Downs recovery plot shown in
  Figure~\ref{f:HD-NG} have error bars that are much larger than one
  would expect for the reported $3-4\sigma$ significance?  \ei

\noindent
Answer: 
While the match for any single angular-separation bin is 
only at the 1-$\sigma$ level (as it should be for 
1-$\sigma$ error bars),~the significance comes 
from the agreement between the measured and expected 
correlations combined over {\it all} pulsar-pair correlations.
In other words, one needs to compute a single 
``signal-to-noise ratio'' by summing inner products of 
the observed pulsar-pair correlations with the expected
Hellings and Downs ``waveform".
(This is similar to what LIGO does when it compares snippets 
of its noisy time-domain data to noise-free chirp templates.)
The individual terms of this sum are inversely weighted by estimates of their
expected squared values {\it in the absence} of the Hellings and Downs
correlations.  So, even though the match at the individual angular
separations are not very close, the overall agreement is quite good.  

\setcounter{footnote}{10} 

In fact, NANOGrav obtains significance estimates%
\footnote{Although 
we focus on NANOGrav, the general discussion applies to all     
PTAs apart from CPTA, which uses a different detection statistic.} 
from two types of analyses:
 
(i) A $3.8 \sigma$ significance determined using frequentist {\it
detection statistics}, such as the noise-weighted inner product of
the observed correlations with the expected Hellings
and Downs correlation, as described above~\cite{PhysRevD.98.044003}.
This is the
``stochastic equivalent" of the deterministic {\it matched filter
 statistic} used by LIGO to search for individual compact binary coalescence
``chirp" signals.  The distribution that would be expected in the
absence of a GWB is estimated by artificially introducing phase
shifts into the timing residual data for each pulsar, 
which removes the Hellings and Downs correlations (see the
right panel of Figure~3 in \cite{NG-GWB:2023}).  The
observed value of the detection statistic lies in the
``tail" of the null distribution, and thus corresponds to high
significance.

(ii) A $3.0 \sigma$ significance obtained via a Bayesian analysis.
This analysis compares the evidence for a data model containing the 
Hellings and Downs spatial correlations to a model without these correlations.
The ratio of the evidences is 226, which
is the ``betting odds" for which of the two models is correct. 
The significance of the value 226 is then estimated from the null 
distribution of the evidence ratio, again constructed using phase
shifts (see both Figure~2 and the left panel of Figure~3 in
\cite{NG-GWB:2023}).
So, what is shown in Figure~\ref{f:HD-NG} for the NANOGrav 15-year
analysis~\cite{NG-GWB:2023} should be thought of as a 
``self-consistency check" of the pulsar-pair cross-correlations, 
{\it calculated under
  the hypothesis that a GWB with Hellings and Downs correlations
matching the Hellings and Downs curve is
present in the data}, with an overall amplitude estimated from
auto-correlation measurements, and assuming that there are no unknown
sources of noise with similar spatial correlations.  The estimated
correlations and error bars are optimally weighted combinations of
approximately 150 correlation measurements in each angular separation
bin, taking into account the covariances between the correlations induced by
the GWB.  If the observed data are consistent with the Hellings and
Downs correlation model, then these point estimates and 1-$\sigma$ error
bars will cross the Hellings and Downs curve approximately
68\% of the time on average.  See~\cite{AllenRomanoOptHD} for a more
detailed discussion.

\vspace{0.1in}\hrule
\bi
\item[Q10:\hspace{-0.05in}] If deviations from the Hellings and Downs curve are detected,
what does that mean? 
\ei

\noindent
Answer: The reconstruction of the Hellings and Downs correlation is
necessarily imperfect, for several reasons.  First, because in
practice, the number of pulsar pairs is finite, whereas the Hellings
and Downs curve is obtained by averaging over an infinite number of
pulsar pairs separated by angle $\gamma$.  Second, because pulsars are
not perfect clocks, and their pulse arrival times are modified by
effects that cannot be completely modeled and 
removed~\cite[Sec.~4]{Lorimer:2008}.
However, if the
number of pulsar pairs is large enough \cite[Fig.~6]{AllenRomanoOptHD} and if
the duration of the observations is long enough, then both of these
issues can be overcome.

So, what if deviations from the Hellings and Downs curve are found,
when there is enough high-quality data from enough pulsars, so that 
neither of these practical issues can be responsible?  In this case,
we have to ask about the size of the deviations, and about how they
are correlated as a function of the angle $\gamma$.  If the deviations
are consistent with the predictions of cosmic variance and covariance
\cite{AllenHDVariance,AllenRomanoOptHD}, it suggests that the GWs
arise from a large number of interfering sources, consistent with a
central-limit-theorem type of process.  Indeed, an {\em exact match} 
to the Hellings and Downs curve would suggest the contrary: that the
GWs are being produced by a small number of GW sources, radiating at 
distinct non-interfering frequencies.  See \cite[Sec.~VIII]{AllenRomanoOptHD} and
\cite[Conclusion]{AllenHDVariance} for further discussion of this
point.


\section{Summary}
\label{s:summary}

We calculated the effect of GWs on the travel time of
electromagnetic signals such as light or radio waves, and showed how the Hellings and Downs curve arises when
correlating timing residuals from PTA pulsars.  We then answered a
number of frequently asked questions (FAQs) about that curve.

Many of these questions arise from inadvertently applying intuition
about the effects of GWs on Earth-based detectors (LIGO, Virgo, Kagra)
to the case of PTAs, where some of that intuition is no longer valid.
The key difference is that a PTA detector has ``arm lengths"
(Earth-pulsar distances) much longer than the wavelengths of the GWs
that they are sensitive to, whereas Earth-based laser interferometers
operate in the opposite (i.e., ``short-arm")
limit.

Calculating the exact response of a single Earth-pulsar baseline to a
passing GW, and comparing limiting expressions of this response for
``short-arm" LIGO-like detectors and ``long-arm" PTA pulsar detectors,
provides qualitative and quantitative answers to many FAQs.  We hope
that this ``Hellings-and-Downs correlation FAQ sheet" will help to
inform the scientific community about the current and ongoing PTA
searches for GWs.

\ack
This paper and its ``FAQ sheet" format arose from discussions that
JDR had with his wife and with several students attending the
``MaNiTou Summer School on GWs," held from 3 July to 8 July 2023, in
Nice, France.  JDR acknowledges past conversations with numerous
NANOGrav colleagues about various aspects of these topics, and
financial support from the NSF Physics Frontier Center Award
PFC-2020265.  He also thanks colleagues Nelson Christensen and
Marie-Anne Bizouard for their support and hospitality while visiting
the Artemis Lab, Observatoire Cote D'Azur in June and July 2023,
where the initial drafts of this paper were written.
The authors thank the referees for helpful comments and 
suggestions, in particular, regarding the close analogy
between the short and long-arm antenna patterns for GW and 
electromagnetic detectors / antennae.

\section*{References}
\bibliography{references}

\end{document}